 \documentclass[final,1p,times,authoryear]{elsarticle}

\usepackage{amssymb}
\usepackage{amsmath}
\usepackage{amsthm}
\newtheorem{theorem}{Proposition}[section] 
\newtheorem{corollary}{Corollary}[theorem]
\newtheorem*{remark}{Remark}
\usepackage[utf8]{inputenc}
\usepackage{todonotes}
\usepackage{url}
\usepackage{natbib}

\journal{Computational statistics and data analysis}
\usepackage{xcolor}

\begin{document}

\begin{frontmatter}



\title{An integrated method for clustering and association network inference} 

 \author[label1]{Jeanne Tous}
 \affiliation[label1]{organization={UMR MIA Paris-Saclay},
             addressline={Universite Paris-Saclay, AgroParisTech, INRAE},
             city={Palaiseau},
             postcode={91120},
             country={France}}

\author[label1]{Julien Chiquet}

\begin{abstract}
High dimensional Gaussian graphical models provide a rigorous framework to describe a network of statistical dependencies between entities, such as genes in genomic regulation studies or species in ecology. Penalized methods, including the standard Graphical-Lasso, are well-known approaches to infer the parameters of these models. As the number of variables in the model (of entities in the network) grow, the network inference and interpretation become more complex. The Normal-Block model is introduced, a new model that clusters variables and consider a network at the cluster level. Normal-Block both adds structure to the network and reduces its size. The approach builds on Graphical-Lasso to add a penalty on the network's edges and limit the detection of spurious dependencies. A zero-inflated version of the model is also proposed to account for real-world data properties. For the inference procedure, two approaches are introduced, a straightforward method based on state-of-the-art approaches and an original, more rigorous method that simultaneously infers the clustering of variables and the association network between clusters, using a penalized variational Expectation-Maximization approach.  An implementation of the model in R, in a package called \textbf{normalblockr}, is available on github\footnote{\url{https://github.com/jeannetous/normalblockr}}. The results of the models in terms of clustering and network inference are presented, using both simulated data and various types of real-world data (proteomics and words occurrences on webpages). 
\end{abstract}


\end{frontmatter}



\section{Introduction}

In statistics, association networks commonly refer to networks used to describe dependency structures between entities. These entities are represented as nodes, and an edge drawn between two nodes indicates a dependency, whose precise meaning is context-dependent. They can be used in psychological science \citep{borsboom2021network} or to represent regulation systems in genomics \citep{fiers2018mapping, lingjaerde2021tailored}, bacterial associations in biology \citep{loftus2021bacterial} or species associations in ecology \citep{ohlmann2018mapping}.  They can be both very informative and complex to analyse as the number of nodes and edges they are made of grows. 

 Undirected graphical models \citep{lauritzen1996graphical, koller2007graphical, whittaker2009graphical} are a convenient and rigorous class of models to represent such networks. In this framework, two nodes are joined by an edge if and only if the random variables they represent are conditionally dependent, given all other nodes of the network. Gaussian graphical models (GGM) enter into this framework. They consider a multivariate Gaussian vector so that partial correlations, and thus the association network, are given by the vectors' precision matrix (the inverse of the variance-covariance matrix). Therefore, the network is described by the model's precision matrix, and its structure corresponds to the support of that matrix. 

 In practice, a GGM's parameters are typically not directly observed. Instead, they need to be estimated from multiple observations of the multivariate Gaussian vector the model describes. Methods have been developed to infer a sparse network from such observations, so as to select the most meaningful edges in the network. A first category of methods consists in multiple testing, to test the presence of each edge individually \citep{drton2007multiple}. The most widespread approaches are penalized methods \citep{yuan2007model, banerjee2008model}: they consist in applying a penalty, often an $\ell_1$ penalty, to the non-diagonal elements of the precision matrix. This leads some of the non-diagonal terms of the precision matrix to be estimated as zero, which translates into the absence of an edge in the graph. This method thus allows to avoid detecting spurious associations (that is, unestablished dependencies).  Graphical-Lasso \citep{friedman2008glasso} is the most popular implementation of the $\ell_1$-penalized approach. Methods with different penalties also exist \citep{chiong2018estimation} as well as more recent approaches that make use of neural networks \citep{belilovsky2017learning} but Graphical-Lasso remains the main reference for inference of sparse networks in GGM. These approaches can also be extended to non-Gaussian data \citep{variational_inf_sparse_network_count_data, liang2023sparse} but in this paper, we will stick to the Gaussian framework.

 As the number of analysed entities grow, the resulting networks become increasingly hard to infer, requiring more data. Large networks are also more complex to analyse both from a computational point of view and for the interpretation of the results. Metrics exist to aggregate information over the whole network such as connectance, nestedness or associations strength \citep{soares2017can, lau2017ecological}. However, such metrics offer very low-grain analysis compared to the complexity of the initial objects they are extracted from. Moreover, they only offer \textit{a posteriori} solutions for the network analysis but they do not reduce the computational cost that comes with Graphical-Lasso inference \citep{mazumder2012graphical}. Nor do they address the fact, that Graphical-Lasso does not make any \textit{a priori} hypothesis on the network structure, even though real networks are usually not Erdös-Rényi graphs, that is edges do not all have the same probability to appear in the network. 

 In order to overcome the computational cost of network inference, \cite{meinshausen2006high} proposed to split the inference into several sub-tasks, using Lasso to infer the neighbourhood of each node in the graph. However, it does not retrieve the variables' individual variance so that one cannot use it to completely retrieve the GGM parameters. \cite{tan2015cluster} consider Graphical-Lasso as a two-step process -- inference of connected components within the graph, and maximization of a penalized log-likelihood on each connected component -- and build on this view to propose another version of the Graphical-Lasso that adds some structure in the network through clustering of the variables and reduces the computational cost by applying the Graphical-Lasso separately in each cluster. 

 Other approaches aim at addressing the issue of the absence of hypothesis on the network structure by identifying or imposing patterns. This can be useful to facilitate network inference, make more hypotheses on its structure and drive the result's interpretation. One can use prior knowledge on the network so as to guide its inference, for instance by forbidding some associations to appear \citep{grechkin2015pathway}. Another method is to assume an underlying structure in the graph, for instance supposing that some nodes display "hub" roles in the network \citep{tan2014learning}, or that edges mostly appear within clusters \citep{ambroise2009inferring}. 
 
 \cite{sanou2022inference} propose an approach based on \cite{meinshausen2006high} and fused-lasso \citep{Pelckmans2005Convex, Hocking2011Clusterpath, Lindsten2011Clustering} that estimates a hierarchical clustering on the variables and a network structure between the clusters. Since it uses the approach of \cite{meinshausen2006high}, this model does not retrieve individual variables' variances. Moreover, the fused-lasso method encourages the inference procedure to retrieve similar dependencies for the elements of the same cluster but it does not directly reduce the size of the network.

 Here, we propose the novel Normal-Block model: a Gaussian Graphical Model with a latent clustering structure on the variables and a network defined at the scale of these clusters. As in other approaches \citep{ambroise2009inferring, tan2015cluster}, we assume that the variables belong to hidden clusters that influence the network structure. The novelty of our method is that we consider a network whose nodes are the clusters (and not the variables themselves). This reduces the dimension of the network so as to simplify both the network's analysis and its inference. The model can also account for the effect of external covariates. For the inference, a first approach consists in using existing methods. To do so, we first use a multivariate Gaussian model on the data. The resulting precision matrix gives a network at the variables level. A clustering of the variables can be done based on the model's residuals. Finally a network at cluster level can be built based on the variables-level network and the clustering. We propose a more ambitious approach that simultaneously clusters the variables and infers the network between the clusters. To this end we resort to variational expectation-maximization to optimize a penalized expected lower-bound of the likelihood. This allows the clustering and the network inference to mutually provide information about one another. We also provide theoretical guarantees on the model's identifiability and inference procedure. Finally, we offer to extend the model to zero-inflated data. 

 We introduce the Normal-Block model in Section~\ref{nb_model} and the corresponding inference strategy in Section~\ref{inference_strategy}. In Section~\ref{zero_inflation}, we show how the model can be extended to zero-inflated data. In Section~\ref{simulation_study}, we study the results we obtain with simulations. Finally, we illustrate the results of the model and its variants (with and without sparsity or zero-inflation) on real-world data  with applications to proteomics data, words occurrences data on web pages and to animal microbiological species in Section~\ref{illustration}.


\paragraph{Notations} Throughout the paper, $\odot$ shall denote the Hadamard product, $\otimes$ the Kronecker product. For a matrix $A$ and an integer $b$, $A^b$ shall denote the matrix with same dimensions as $A$ obtained by individually raising each element of $A$ to the power of $b$, and $A^{\oslash}$ the term-to-term inverse of $A$, that is  $A^{\oslash} = (A^{-1}_{ij})_{i \in [\![1;n]\!], j \in [\![1;p]\!]}$. Similarly $f(A)$ will correspond to the term-to-term application of function $f$ to matrix $A$, that is $f(A) = (f(A_{ij}))_{i \in [\![1;n]\!], j \in [\![1;p]\!]}$. $A_{i.}$ denotes the $i$-th row of matrix $A$ and $A_{.j}$ its $j$-th column, whereas $A_{i}$ (no $.$) denotes the transposed $i$-th row of matrix $A$ , a column vector. We use  $A_{row-sum}$  $= (\sum_i A_{i.})^T$, $A_{col-sum}$ $= \sum_j A_{.j}$ and $A_{total-sum}$ $= \sum_{ij} A_{ij}$
\section{An integrated model of clustering and network reconstruction for continuous data} \label{nb_model}
\subsection{The Normal-Block model}

We observe $\{Y_i, 1 \leq i \leq n\}$, $n$ realizations of a $p$-dimensional Gaussian vector so that $Y_i$ may describe the expression intensities of $p$ genes in cell $i$ or the biomass of $p$ species in site $i$.

The model relates each continuous vector $Y_i \in \mathbb{R}^p$ ($1 \leq i \leq n$) to a vector of latent variables of smaller dimension $W_i \in \mathbb{R}^q, q < p$, with precision matrix $\Omega$ (that is, covariance matrix $\Sigma = \Omega^{-1}$). As in a multivariate linear model, we also include the effects of a combination of covariates $X_i \in \mathbb{R}^d$, with $d \times p$ matrix $B$, the matrix of regression coefficients. 
\begin{equation}
    \label{eq:nb-model}
    \begin{aligned}
        & \text{Latent space: } W_i \sim \mathcal{N}(0, \Omega^{-1})\\
        & \text{Observation space: } Y_i \ | \ W_i \sim  \mathcal{N}(CW_i + B^{\top}X_i , D)
    \end{aligned}
\end{equation}

We denote the observed matrices by $Y$ and $X$, with sizes $n \times p$, $n \times d$ stacking vectors row-wise, and $W$ the $n \times q$ matrix of latent Gaussian vectors. The $p \times q$ matrix $C$ is a clustering matrix with $C_{jk} = 1$ if and only if entity $j$ belongs to cluster $k$. This clustering links observations $Y_i$ and latent variables $W_i$. $C$ can either be observed or not. When $C$ is observed, the framework of Model~\eqref{eq:nb-model} is that of multivariate mixed models, with $B$ a matrix of fixed effects with design matrix $X$, and $W$ a matrix of random effects with $C$ being the corresponding design matrix. When $C$ is unobserved, we further assume that the $j$-th column of $C$, denoted $C_j\in\{0,1\}^{q}$ is a multinomial random variable, that is: $C_j \sim \mathcal{M}(1, (\alpha_{jk})_{1 \leq k \leq q})$ so that $\sum_{k=1}^{q} \alpha_{jk} = 1$. 

Model~\eqref{eq:nb-model} relates observations $Y_i$ both to observed covariates $X$ and to a clustering effect that translates into $W_i$'s covariance. The addition of a diagonal variance matrix $D$ in the conditional distribution of $Y$ aims at separating the effect of variables' individual variance to ensure that the clusters' effects on covariances is not biased by individual variance effects. We can also consider a spherical model, forcing individual variances to be the same for each entity so that $D$ becomes a spherical variance matrix: $D = \mathrm{diag}(\xi^{-1}), \xi \in \mathbb{R}^{+*}$. The set of model parameters is denoted as $\theta = (B, \Omega, D)$.

 This framework allows the modelling of small networks (of size $q \times q$) from large datasets, based on a  clustering of the entities, as we detail in Section \ref{graphical_model}.

\subsection{Graphical model}\label{graphical_model}
The goal of the model is to consider a clustering (whether it is observed or not) of continuous variables  and an association network between the clusters. To do so, we resort to the framework of graphical models \citep{lauritzen1996graphical}. The association network encodes the dependencies between the components of the latent variable $W$, corresponding to the observations $Y$'s residuals after accounting for covariates. More precisely, variables $W_{k_1}$ and $W_{k_2}$ are connected in the graph if they remain dependent after conditioning on all other $W_l$. Since the $W$ are jointly Gaussian, this dependence corresponds to a non-zero value in the precision matrix $\Omega$ of the Gaussian distribution, that is: $W_{k_1}$ and $W_{k_2}$ are connected in the network if and only if $\Omega_{k_1 k_2} \neq 0$. The partial correlation between them is then given by $-\Omega_{k_1 k_2} /\sqrt{\Omega_{k_1 k_1}\Omega_{k_2 k_2}}$. Thus, the association network between the $q$ clusters is represented in the dependency structure between the components of the latent variable $W_i$ from one site $i$ to another, and it is encoded in $W_i$'s precision matrix $\Omega$ of size $q \times q$.

 The structure of the network is determined by the support of $\Omega$. To limit the detection of spurious associations, we may want to infer a sparse network. To do so, we add an $\ell_1$ penalty on $\Omega$ in the likelihood or its variational approximation in the inference procedure. We resort to Graphical-Lasso to implement this regularization \citep{friedman2008glasso}.

\subsection{Comparison with the factor analysis}

In its writing and, to a certain extent, in its philosophy, the Normal-Block model is similar to the well-known factor analysis \citep{tipping1999probabilistic, murphy2022probabilistic}. As described by \cite{murphy2022probabilistic}, factor analysis can be seen as a "low-rank version of a Gaussian distribution". It can be written as a latent variable model:

\begin{equation}
    \label{eq:factor-analysis}
    \begin{aligned}
        & \text{Latent space: } W_i \sim \mathcal{N}(\mu_0, \Sigma_0),\\
        & \text{Observation space: } Y_i \ | \ W_i \sim  \mathcal{N}(CW_i + \mu, D),
    \end{aligned}
\end{equation}

with $Y_i$ of dimension $p$, $W_i$ of dimension $q < p $, $C$ a $p \times q$ matrix called the \textit{factor loading matrix} and $D$ a $p \times p$  diagonal matrix. As the effects of $\mu_0$ can be absorbed into $\mu$, one can set $\mu_0 = 0$ without loss of generality. Replacing $\mu$ with a covariate effect $B^{\top}X_i$ as is done in the Normal-Block model would be a mild modification of the factor analysis model and would not fundamentally change it. As explained by \cite{murphy2022probabilistic}, in this model, $C$ can also be replaced by $\Tilde{C} = C\Sigma_0^{-1/2}$ so that, without loss of generality, one can set $\Sigma_0 = I_q$, the $q \times q$ identity matrix.

In their writing, the main difference between the two models are that the Normal-Block "factor loading matrix", $C$, is a \emph{clustering matrix} that cannot be modified to replace $\Sigma$ with $I_q$.

Both models use lower dimensions latent variables to consider lower number of parameters. However, factor analysis considers each observation as a combination of several, lower-dimensions effects, with an additional noise. Its goal is to find uncorrelated underlying axes that help analyse the observations. The Normal-Block model takes a different approach in the sense that it aims at relating one observation with a single underlying lower-dimension variable through clustering (in $C$). This also explains why $C$ cannot be modified, as in the factor analysis, to replace $\Omega^{-1}$ with $I_q$. The structure it considers is that of variance-covariance between the latent variables. The main aim of the model is to identify an underlying correlation structure in the variables. This is why the Normal-Block model is also related to the Gaussian graphical model framework, as explained in section \ref{graphical_model}, whereas that is not the idea of factor analysis.
In the Normal-Block framework, when $C$ is observed and $\Omega = I_q$, one considers that the clusters are uncorrelated and that the underlying association network is empty of edges. In this special limiting case, the Normal-Block model amounts to a factor analysis. One can see that the EM strategy to estimate the Normal-Block model parameters when $C$ is observed is similar to that described by \cite{murphy2022probabilistic} for factor analysis.

The factor analysis model is not identifiable because any orthogonal rotation of $C$ yields the same likelihood. This issue can be overcome by adding constraints on $C$ (forcing its column to be orthogonal as in PCA, or forcing it to be lower triangular for instance, see \cite{murphy2022probabilistic}). The fact that, in the Normal-Block model, $C$ is constrained to be a clustering matrix is also what allows one to prove its identifiability. \\

\subsection{Identifiability}\label{identifiability}
In this section, we prove that the Normal-Block models, both with observed and unknown clusters, are identifiable under mild conditions. 
\subsubsection{Observed clusters model}

\begin{theorem}
The spherical Normal-Block model with observed clusters is identifiable provided $X$ has rank $d$, no cluster is empty and at least one cluster contains at least two elements.
\end{theorem}

\begin{theorem}
The Normal-Block model with observed clusters is identifiable provided $X$ has rank $d$ and each cluster contains at least two elements.
\end{theorem}

The proofs for both propositions are presented in Appendix \ref{appendix_identifiability}. 

\begin{remark}
These propositions express the intuitive idea that the variables' variances is a mix of their cluster's variance and their individual variance. If the group is only made of one element then these two variances represent the same thing. In the spherical model case the hypothesis is less constraining because the individual variances are assumed to be the same for all. 
\end{remark}

\begin{remark}
 Should the hypothesis on the number of elements in each cluster not be respected, the model's interpretation would not be hindered. Indeed if cluster $k$ contains only category $j$, one would simply need to consider the sum $\Sigma_{k k} + D_{jj}$ as both the cluster and individual variance. This makes sense as, in that case, both correspond to the same entity. 
 \end{remark}

\subsubsection{Unknown clusters model}

When the clusters are unobserved, the identifiability becomes more complex to prove. The marginal likelihood is
\begin{align*}
    p_{\theta}(Y_i) = \sum_{C^* \in [\![1; q ]\!]^p} \left(\Pi_{j=1}^p \alpha_{C_j^*}\right) \mathcal{N}(Y_i | B^{\top}X_i, D + C^* \Sigma C^{* \top}).
\end{align*}

\begin{theorem}
The Normal-Block model with unknown clusters is identifiable provided $p > q$, $X$ has rank $d$, for all $k \in [\![1 ; q]\!], \alpha_k > 0$, the diagonal values of $\Sigma$ are two-by-two distinct, and no non-diagonal value of $\Sigma$ is equal to one of its diagonal values.
\end{theorem}

The proof for this proposition is detailed in Appendix \ref{appendix_identifiability}. It is based on the identifiability of a finite mixture of Gaussian distributions given by \cite{yakowitz1968identifiability}.

\section{Inference strategy} \label{inference_strategy}

We now describe the inference strategies, that aim at estimating $B, \Omega, D$ (or $\xi$ in the spherical model), and, when not observed, $C$. We first introduce a 2-step approach (\ref{first_inference_method}) based on existing methods from the literature, before proposing a more rigorous, fully integrated approach (\ref{EM_observed}, \ref{EM_unobserved}) relying on Expectation-Maximization (EM) when the clusters are observed and on Variational EM when they are not, simultaneously inferring the clustering and the network. Finally, we discuss model selection in Section \ref{model_selection}.

\subsection{An inference approach based on state-of-the art methods}\label{first_inference_method}

We first study how state-of-the art methods could be used to infer in part the model's parameters and how well they would perform. Combining the Graphical-Lasso \citep{friedman2008glasso} for the GGM side and SBM \citep{holland_sbm, chiquetsbm} or k-means for the clustering side leads to a 2-step procedure described herein. Consider the Gaussian multivariate linear model 
\begin{equation}
    \label{eq:mvn-model}
    Y_i = B^\top X_i+ R_i, \text{ with } R_i \sim \mathcal{N}(0, \Gamma),
\end{equation} 
with $\Gamma$ a $p\times p$ covariance matrix. For $B$ and $\Sigma$, we use the standard multivariate linear regression estimators: $\hat{B} = (X^{\top}X)^{-1}X^{\top}Y$, $\hat{R} = Y - X\hat{B}$ and $\hat{\Gamma} = \hat{R}^{\top}\hat{R} / n$. 

Then, if the clustering $C$ is not observed, we either estimate it with a k-means algorithm on $\hat{R}$, or with a stochastic block model (SBM) \citep{holland_sbm, chiquetsbm} on $\hat{\Gamma}$. Once we have a clustering $C$ (observed or estimated), the empirical $q\times q$ covariance $\tilde{\Sigma}$ between groups is estimated with
\begin{align*}
    \tilde{\Sigma}_{k_1, k_2} &= \frac{1}{|k_1| \times |k_2|}\sum_{j_1, j_2, C_{j_1 k_1}C_{j_2 k_2} = 1}\hat{\Gamma}_{j_1,j_2 } ,
\end{align*}
where $|k_1|, |k_2|$ are the number of elements in clusters $k_1$ and  $k_2$ respectively. This means that the element of $\tilde{\Sigma}$ that corresponds to the covariance between clusters $k_1$ and $k_2$ is estimated from the covariance terms of $\hat{\Sigma}$ that concern each pair of elements with one element in cluster $k_1$ and the other in cluster $k_2$, weighted by the number of elements in each of these clusters.

Finally, the sparse estimator of $\Sigma$ is obtained by applying Graphical-Lasso to $\tilde{\Sigma}$. We use the implementation provided in the package \textbf{glassoFast} \citep{sustik2012glassofast}. \\
However, with this method, the effects of $\Sigma$ and $D$ cannot be properly teased apart.

The primary benefit of this method is its simplicity and intuitiveness. Additionally, its outcomes can serve as an initial input for our more elaborate, integrated inference approach. However, the 2-step method lacks rigorous justification and cannot properly estimate all model parameters because it does not optimize a specific criterion like the likelihood. This limitation makes it difficult to assess the model's fit to the data and prevents the development of a statistically sound criterion for selecting the number of groups or edges in the network. Furthermore, using existing approaches to this problem implies inferring the network and clustering separately. The network is first retrieved at the variable level, which does not reduce dimensionality. Additionally, handling clustering separately prevents the two processes from informing each other. Therefore, in the subsequent sections, we propose a more rigorous approach that simultaneously infers both the clustering and the network at the cluster level.

\subsection{Expectation-Maximization method for the observed clusters model}\label{EM_observed}

We now introduce a fully integrated likelihood-based approach to infer the model parameters $\theta = (B,\Omega, D)$. We first consider the inference when the clustering is observed.\\

If $C$ is given and fixed, we can write the marginal distribution of $Y_i$:
\begin{equation*}
    Y_i \sim \mathcal{N}(B^T X_i, D + C \Omega C^{\top}),
\end{equation*}
However, the marginal likelihood does not allow one to tease apart the effects of $D$ and $C \Omega C^{\top}$ in the variance. Thus we instead resort to the complete likelihood $\log p (Y_i, W_i)$, using an Expectation Maximization (EM) strategy \citep{dempster1977maximum}.

The E step consists in evaluating the conditional expectation $\mathbb{E}_{W_i\sim p(. | Y; \theta')}[\log p_\theta (Y_i, W_i)]$, which requires the characterization of the posterior distribution $W_i \ | \ Y_i$. Since $W_i$ and $Y_i$ are both Gaussian variables, we can explicitly develop the expression of the conditional density using Bayes formula and we get
\begin{equation*}
    W_i \ | \ Y_i \sim  \mathcal{N}(\mu_i, \Gamma), \quad \text{with }
    \Gamma = (C^\top D^{-1}C + \Omega)^{-1} \ \text{and} \ \mu_i = \Gamma C^\top D^{-1} (Y_i - B^{\top}X_i).
\end{equation*}

Then, we easily derive the expression for the EM criterion, which serves as our objective function in $\theta$ from an optimization standpoint:
\begin{align*}
    J(\theta) & = -\frac{np}{2} \log(2 \pi)  - \frac n2 \log\det(D) - \frac 12\mathrm{ tr}\left(D^{-1} \left(R_\mu^\top  R_\mu + n C \Gamma C^{\top}\right)\right)\\
     & \quad - \frac{nq}{2}\log(2 \pi) + \frac n2 \log\det(\Omega) - \frac 12\mathrm{ tr}\left(\Omega \left(n\Gamma + \mu^\top \mu\right)\right)
\end{align*}
with $R_\mu = Y - XB - \mu C^{\top}$  and $\mu$ the matrix whose rows are the $\mu_{i}^\top$, that is $\mu^\top = \Gamma C^{T}D^{-1}(Y - B^\top X)$. \\

The M-step consists in updating the parameters by maximizing $J$ with respect to (w.r.t.) $\theta$. Closed-forms are obtained for the different parameters $(B,\Omega,D)$  by differentiation of the objective function $J(\theta)$, so that the concavity, at least in each parameter, is needed. This is stated in the following Proposition, the proof of which is presented in Appendix \ref{appendix_concavity}.
\begin{theorem}
\label{prop:concavity}
In the observed-clusters model, the objective function $J$ is jointly concave in $(\Omega, D^{-1})$ and in $(\Omega, B)$. The same holds for the spherical model, with joint concavity in $(\Omega, \xi^{-1})$ and in $(\Omega, B)$.
\end{theorem}

We now state the closed-form expressions for the estimators calculated during the M-step:
\begin{theorem}\label{estimators_observed_clusters}
The M-estimators for the observed-clusters model, obtained by differentiation of $J$ w.r.t. $B$, $\Sigma$, $d$ the diagonal vector of $D$ and $\xi$ for the spherical model are given by
\begin{align*}
    \hat{B} &= \left(X^T X\right)^{-1} X^T (Y - \mu C^T), \quad
    \hat{\Sigma} = \frac 1 n \mu^T\mu + \Gamma, \\
    \hat{d} &= (R_\mu^2)_{row-sum}/n + C\mathrm{diag}(\Gamma), \quad
    \hat{\xi}^2 = R_{\mu^2_{total-sum}}/{np} + C_{row-sum}^T\mathrm{diag}(\Gamma)/p.
\end{align*}
 \end{theorem}

As mentioned in Section~\ref{graphical_model}, we may also want to add an $\ell _1$ penalty on $\Omega$ so as to infer a sparse network structure. In that case, the objective function becomes $J_{\text{struct}} = J - \lambda ||\Omega||_{\ell_1, \text{off}}$, where $||\Omega||_{\ell_1, \text{off}}$ is the off-diagonal $\ell_1$ norm of $\Omega$ and $\lambda > 0$ is a tuning parameter to control the sparsity level. We only penalize the off-diagonal element of $\Omega$ since we only want to restrict the associations, not the intra-clusters variances. $J_{\text{struct}}$ is a lower bound of $J$, whether the clusters are observed or not. Thanks to the concavity of $\Omega \mapsto - \lambda ||\Omega||_{\ell_1, \text{off}} $, adding the penalty does not change the structure of the objective function, which preserves its concavity property:

\setcounter{corollary}{0}
\addtocounter{theorem}{-1}
\begin{corollary}
\label{prop:concavity_struct}
In the observed-clusters diagonal and spherical models, the penalized objective function $J_{\text{struct}}$ is jointly concave in $(\Omega, D^{-1})$ and in $(\Omega, B)$.
\end{corollary}

The E-step and  M-step are similar in the sparse case: we evaluate $J_{struct}(\theta)$ and estimate the model parameters as stated in Theorem~\ref{estimators_observed_clusters}. The only striking difference lies in the estimation of $\Omega$, obtained here with Graphical-Lasso applied to $\hat{\Sigma}$. We use the implementation provided in the package \textbf{glassoFast} \citep{sustik2012glassofast}. \\

The whole EM algorithm is initialized with the 2-step method described in Section~\ref{first_inference_method}, from which we obtain starting values for the parameters $B$ and $\Sigma$. The rest of the optimization then consists in alternately updating the parameters of the posterior distribution ($\Gamma, \mu$)  in the E-step and, and of the model parameters ($B, D, \Omega)$ in the M-step.
 
\subsection{Variational inference for the unobserved clusters method}\label{EM_unobserved}

 When the clustering $C$ is not observed, we can also propose an integrated inference method, allowing the clustering and the network at cluster level to simultaneously be inferred. We assume that the number of clusters $q$ is fixed as a hyper-parameter. The marginal likelihood can no longer be computed so that we also need to resort to an EM strategy. However, this requires computing some moments of $W, C \ | \ Y$ since they are required in the E step for the evaluation of $\mathbb{E}_{C, W_i \sim p(. | Y; \theta')}[\log p_\theta (Y_i, W_i, C)]$. Since these posterior distribution moments are untractable, we resort to a variational approximation \citep{blei2017variational, wainwright2008graphical} and proceed with a Variational-EM (VEM) algorithm.

\subsubsection{Variational approximation}
Under the variational approximation, we assume that:
\begin{align*}
    \mathbb{P}(W, C | Y) & \sim  \mathbb{P}(W | Y) \mathbb{P}(C | Y) \\
    &  \sim \Pi_{i=1}^n \mathbb{P}(W_i | Y)  \Pi_{j=1}^p \mathbb{P}(C_j | Y)  \\
    &  \sim \Pi_{i=1}^n \pi_1(W_i)  \Pi_{j=1}^p \pi_2(C_j) 
\end{align*}
with:
\begin{itemize}
    \item $\pi_1$ the approximation for $W| Y$: $W_i \sim^{\pi_1} \mathcal{N}(M_i, S_i)$, $S_i$ being diagonal. We denote $S \in \mathcal{M}_{n, q}(\mathbb{R})$ defined by $S_{i, k} = s_{i, k}$ and $M \in \mathcal{M}_{n, q}(\mathbb{R})$  defined by $M_{i, k} = M_{i_{k}}$ so that the parameters of $\pi_1$ can be denoted $\psi_1 = (M, S)$.
    \item  $\pi_2$ the approximation for $C| Y$:  $C_j \sim^{\pi_2} \mathcal{M}(1, (\tau_{jk})_{1 \leq k \leq q})$ and $\forall j \in [\![1 ; p ]\!], \sum_{k=1}^{q} \tau_{jk} = 1 $. We denote $\tau \in \mathcal{M}_{p, q}(\mathbb{R})$ the matrix of $(\tau_{jk})_{j \in [\![1;p]\!], k \in [\![1;q]\!]}$ so that the parameters of $\pi_2$ can be denoted $\psi_2 = (\tau)$.
\end{itemize}

 The quality of this approximation is measured with the Kullback-Leibler divergence between the two distributions (that is, the true, untractable $W, C | Y$ and the distributon $\pi_1 \pi_2$). This  allows us to write a variational lower-bound (ELBO or Expected Lower Bound) of the marginal log-likelihood:
\begin{align*}
    J & = \log p_{\theta}(Y) - KL[\pi_1(W)\pi_2(C) | Y || p_{\theta}(W, C | Y)] \\ \\
    &= \mathbb{E}_{\pi}(\log p_{\theta}(Y, W, C)) -  \mathbb{E}_{\pi_1}(\log \pi_1( W)) -  \mathbb{E}_{\pi_2}(\log \pi_2(C)) \\ \\
    & = - \frac n2 \log(2\pi) - \frac{n}{2}\log(\det(D)) - \frac 1 2 \mathbf{1}_n^T \left(A D^{-1} \right)\mathbf{1}_p \\
    & \quad - \frac{nq}{2} \log(2 \pi) + \frac{n}{2}\log\det(\Omega) - \frac 12 \mathrm{tr}\left(\Omega (\mathrm{diag}(S_{row-sum}) + M^{\top}M) \right) \\
    & \quad + \frac{nq}{2}\log(2 \pi e) + \frac{1}{2}1_n^\top\log(S)1_{q} + 1_p^\top\tau\log(\alpha) - 1_p^\top \left((\tau \odot \log(\tau)) \right)1_{q}, 
\end{align*}

\noindent where $R = Y-X B$ and $A=R^2 - 2 R \circ M \tau^T + (M^2+S)\tau^T$, and $S_{row-sum} = \sum_{i=1}^n S_i$.

\subsubsection{Concavity}
\begin{theorem}
In the unobserved-clusters model the objective function $J$ is individually concave in each of its terms. 
\end{theorem}

 A proof of this proposition is presented in  Appendix \ref{appendix_concavity}.  We do not have the global concavity of $J$ so that its convergence towards a global optimum is not guaranteed. 

\begin{corollary}
In the unobserved-clusters model the penalized objective function $J_{\text{struct}}$ is individually concave in each of its terms.
\end{corollary}

 This corollary arises from the  concavity of $\Omega \mapsto - \lambda ||\Omega||_{\ell_1, \text{off}} $.

\subsubsection{Inference algorithm}
 Similarly to the observed clusters case, the VEM algorithm consists in alternately computing estimators for the variational parameters $M, S, \tau$ in the VE-step and for the model parameters $\alpha, B, D, \Omega$ in the M-step.  First-order derivatives of the ELBO give explicit estimators for all of these parameters. The expressions obtained thereby are given in Proposition~\ref{estimators_unobserved_clusters}. 

\begin{theorem}\label{estimators_unobserved_clusters}
In the unobserved-clusters case, estimators of the variational parameters in the E-step are
\begin{align*}
    \hat{M} &= R D^{-1} \tau {\tilde{\Gamma}}, \quad
    \hat{S}_i = \mathrm{diag}( {\tilde\Gamma}), \quad
    \hat{\tau}_j = \mathrm{softmax}(\eta_j),
\end{align*}
where we denote $ {\tilde{\Gamma}} = (\Omega + \mathrm{Diag}(\tau^T d^{-1}))^{-1}$ and $\eta = -\frac 1 2 d^{-1} \otimes (M^2_{row-sum} + n S_i) + D^{-1} R^T M + \mathbf{1}_p\otimes \log(\alpha) - 1$.

For the M-step, we have
\begin{align*}
    \hat{B} &= \left(X^T X\right)^{-1} X^T (Y - M\tau^T),\quad
    \hat{\Sigma}_{q} = \frac 1 n (M^TM + \mathrm{Diag}(S_{row-sum})), \quad
    \hat{d} = \frac 1 n A_{row-sum},
\end{align*}
\noindent denoting $A=R^2 - 2 R \circ M \tau^T + (M^2+S)\tau^T$.
\end{theorem}

For the initialization we also need to specify an initial clustering. We still rely on our 2-step approach where, when the clustering in unobserved, we use the k-means algorithm on the residuals of a multivariate Gaussian model or a SBM on the empirical covariance at the variable scale.

\subsection{Model selection}\label{model_selection}

There are two underlying hyper-parameters to the Normal-Block model. The first one is the penalty $\lambda$ applied on the precision matrix: the higher it gets, the sparser is the resulting network. The second one, when the clustering is not observed, is the number of clusters $q$. While there is no exact method to fix these two parameters, we propose several approaches here.

\subsubsection{Selecting the number of groups}
As is often the case in clustering problems, the number of clusters, $q$ here, is a hyper-parameter. The Bayesian Information Criterion (BIC), the Extended BIC (EBIC) and the Integrated Complete Likelihood (ICL) \citep{biernacki2002assessing} can be used as criteria to fix $q$. While the model's likelihood increases with $q$ because the number of parameters does so, these criteria penalize a too important number of parameters and help find a balance. Empirically, on simulated data, with $n \in \{50, 100, 200\}, p = 100 $ and $q \in \{3, 5, 10\}$, we find that the BIC and the EBIC retrieve the correct number of clusters in more than $99 \%$ cases while the ICL does so in more than $97\%$ of them. When an error is made, the number of clusters identified by the criterion is either equal to $q + 1$ or $q - 1$.

\subsubsection{Selecting the penalty}
Adding a penalty $\lambda$ on $\Omega$ helps infer a sparse network. Several approaches exist to find the optimal $\lambda$ in GGM. A higher $\lambda$ will force more values of $\Omega$ to 0 and hence reduce the number of parameters and the likelihood while statistical criteria such as the BIC, the ICL or the EBIC \citep{chen2008extended} will favour both an increasing likelihood and a lower number of parameters. One can rely on these criteria to fix $\lambda$.

Another approach that we propose is the Stability Approach to Regularization Selection (StARS) \citep{liu2010stability}. In short it consists in recomputing networks from data subsamples for each $\lambda$ and to keep the value of $\lambda$ for which the networks inferred with the different subsamples are the most stable, stability being measured through the frequency of appearance of the edges in the networks obtained from the different subsamples. As the Normal-Block networks are built at the scale of clusters, we need to fix $q$ and the clustering to proceed to StARS. For a fixed $q$, we propose to keep it as it is when no penalty is applied on the network ($\lambda = 0$). StARS then keeps the lowest penalty such that all the edges it identifies are present in more than $x \%$ of the networks obtained from subsamples, $x$ being a hyper-parameter called stability threshold and taking values between $0$ and $1$. 

\begin{figure}
    \centering
    \includegraphics[width = 15cm]{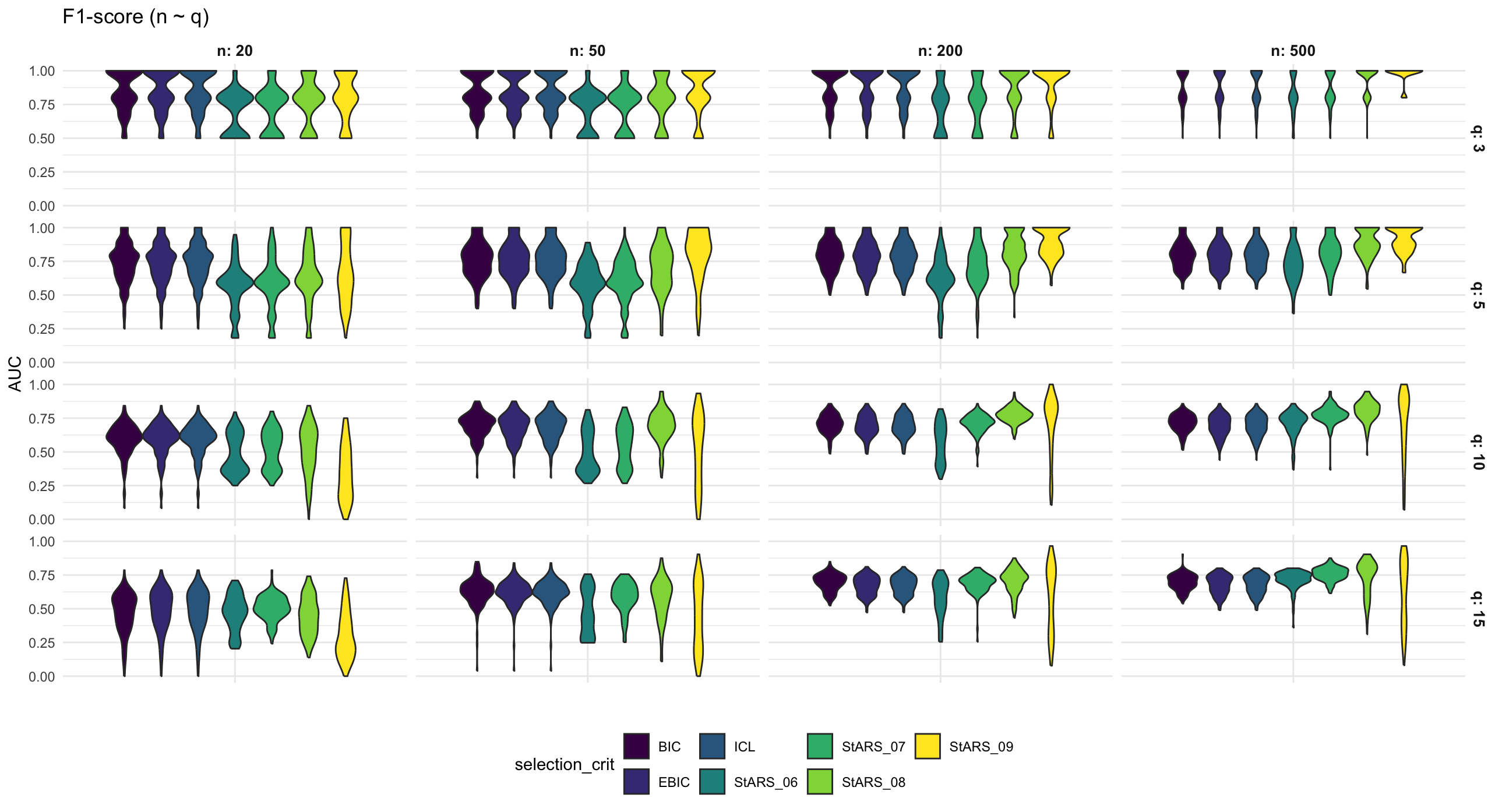}
    \caption{Violin plots of the F1 scores obtained with fixed blocks, for $q \in \{3, 5, 10, 15\}$ and $n \in \{20, 50, 200, 500\}$  for the penalty retrieved using either BIC, EBIC or ICL criteria, or a StARS method with stability fixed at $0.6, 0.7, 0.8, 0.9$.}
    \label{fig:f1_res}
\end{figure}

We compare the criterion-based approaches and the StARS approach using the F1-score, equal to 2  $\times$ (precision $\times$ recall) / (precision + recall) (Figure \ref{fig:f1_res}).

Figure \ref{fig:f1_res} shows that the three criteria obtain similar results in terms of F1 score. StARS results can get much worse when using a stability value that is too high (0.9). For lower stability values, the results are comparable. Overall the BIC, EBIC and ICL criteria seem to offer results comparable to that of StARS and more stable. Moreover, using StARS is computationally expensive due to the resampling procedure so that it seems preferable to resort to one of the statistical criteria.

\section{Extension to zero-inflated Gaussian data} \label{zero_inflation}
Additionally to the model and inference methods we have proposed, we offer to extend the model to zero-inflated data in this section. This can be useful in situations where real-world data display zero-inflated patterns (that is, contains more zeros than can be explained by a Gaussian distribution) whether it is because of technical limitations, variations in sampling efforts or other reasons. This is often the case in ecology due to sampling procedure or in genomics, with single-cell experiments for instance. As in Section \ref{inference_strategy}, we also propose a multiple-step method for parameters inference (\ref{multiple_step_zi}) and an EM-based method (\ref{vem_zi}).

\subsection{Model}
For the zero-inflated version of Normal-Block, taking inspiration from \cite{chiquet2024zero}, we add a second latent variable $Z$, such that each $Z_{ij}$ follows a Bernoulli distribution of parameter $\kappa_j$:
\begin{align*}
    \text{Gaussian latent layer: } & W_i \sim \mathcal{N}(0, \Omega^{-1})\\
    \text{Excess of 0 latent layer: } & Z_i \sim \bigotimes_j \mathcal{B}_j(\kappa_j) \\
    \text{Observation space: } & Y_i | Z_i, W_i = Z_i \odot \delta_0 + (1-Z_{i}) \odot (C W_{i} +  \mathcal{N}(B^\top X_i , D))
\end{align*}

\noindent where $ \delta_0$ is the Dirac distribution in 0.

\subsection{2-step inference approach based on state-of-the art methods}\label{multiple_step_zi}
We propose a 2-step straightforward inference method. We first infer $B$ and $\kappa$ as the parameters of a zero-inflated diagonal Normal model, defined as:
\begin{equation*}
    Y_{i} \, | \, Z_{i} = \delta_0 Z_{i} + (1 - Z_{i}) \mathcal{N}(B^\top X_i, D).
\end{equation*}

 For parameter inference, we resort to an EM approach. Since the normal distribution is a continuous distribution, access to the posterior probability $Z | Y$ is straightforward: $p(Z_{ij} = 1 | Y_{ij}) = 1_{Y_{ij} = 0}$. We can compute residuals $\tilde{R} = Y - XB$ and obtain $\hat{R}$ from $\tilde{R}$ replacing $\tilde{R}_{i,j}$ with 0 when $p(Z_{ij} = 1 | Y_{ij})$, that is when $Y_{ij} = 0$. From here, we compute $\hat{\sigma} = \hat{R}^{\top}\hat{R}/n$. As for the non-zero-inflated data (\ref{first_inference_method}), when the clustering $C$ is not observed, several clustering methods are proposed to infer it, including a k-means algorithm on $R$ or a SBM \citep{holland_sbm, chiquetsbm} on $\hat{\sigma}$, and from then obtain $\hat{\Sigma}$ the same way.

\subsection{EM-based inference method}\label{vem_zi}
For the observed cluster models, one can compute the marginal log-likelihood or the complete log-likelihood without variational approximation as the posterior distribution of $Z_i$ is straightforward ($p(Z_{ij} = 1 | Y_{ij}) = 1_{Y_{ij} = 0}$) and that of $W_i$ is similar to the non-zero-inflated case, after removing the zeros. We can thus proceed to an EM-inference. However when the clustering is unobserved, we need to resort to VEM-inference, as in the non-zero-inflated case. Again, we resort to a mean-field approximation with $W_i | Y_i$ approximated by $\pi_1 $ with $W_i \sim^{\pi_1} \mathcal{N}(M_i, S_i)$, $S_i$ being diagonal. Details are given in Appendix~\ref{appendix_zi_estimators}.

\section{Simulation study}\label{simulation_study}
We study the performance of our inference methods on data simulated under the Normal-Block model, with and without zero-inflation. The code used to simulate the data is available in the \textit{inst} folder
of the \textbf{normalblockr} github repository. We first want to test their ability to retrieve the correct clustering when it is not observed, and the structure of the association network (that is, the support of the association matrix). To assess how the network is retrieved, we only consider observed-clusters simulations. Indeed, when clusters are unobserved, the model is only identifiable up to label permutation and comparing the networks requires testing all these permutations. As the clustering is usually well-retrieved by the integrated inference method, one can imagine that the results would be similar between observed and unobserved clusters simulations for network inference. More generally, we also test the inference methods' ability to retrieve each parameter's value. 

We want to compare these results for different levels of difficulties. We assume that increasing $q$ and decreasing $n$ make the inference harder as it means more information to retrieve with less signal. Similarly, increasing the level of zero-inflation is likely to degrade the results as it removes some of the signal in the data. We also run simulations for different structures of $\Omega$  to see if some network structures are easier to retrieve than others. 

Finally, to test the robustness of the method when the clustering is inaccurate (either because a wrong clustering is given as an input or because the inference method makes a mistake in the clustering), we test the results of the integrated inference when errors are introduced in the clustering. We introduced wrong labels mistakes for $5 \%$ to $15\%$ of the variables.

\subsection{Simulation protocol}
\subsubsection{Network generation}\label{network_generation}
To generate the ground-truth $\Omega$, we first produce a sparse undirected graph with different possible structures: Erdös-Rényi (no particular structure), preferential attachment (edges are attributed progressively with a probability proportional to the number of edges each node is already involved in) and community (in the Stochastic Block Models sense, \cite{holland_sbm}). This allows us to test the robustness of the inference procedure when facing different dependency structures. Using package \textbf{igraph} \citep{csardi2006igraph}, we generate an adjacency matrix $G$ corresponding to a given structure. Then $\Omega$ is created with the same sparsity pattern as $G$, as follows: $\tilde{\Omega} = G \times v$ and $\Omega = \tilde{\Omega} + \mathrm{diag}(\min(\mathrm{eig}(\tilde{\Omega}))| + u)$, with $u, v > 0$ two scalars. Higher $v$ means stronger correlations whereas higher $u$ means better conditioning of $\Omega$. Following \cite{variational_inf_sparse_network_count_data} we fix $v = 0.3, u = 0.4$ in the simulations. 

\subsubsection{Data generation}
We simulate data under the Normal-Block model. We draw a random but balanced clustering (variables are affected to each cluster with equal probability), a unique covariate taking values in $[1 ; 10]$ then draw $Y$ according to the model. For non-zero-inflated Normal-Block we test $n = 20, 50, 200 \text{ or } 500$, $p = 100 \text{ or } 500$ and $q= 3, 5, 10 \text{ or } 15$. 

Inference for zero-inflated data takes longer, especially as $q$ increases so that we only test $q = 3, 5$, $n = 75$ and Erdös-Rényi graph structure, but we test different levels of zero-inflation. $\kappa$ is drawn from a truncated Gaussian distribution, with standard deviation of $\sigma = 0.05$, mean $\mu$ either equal to $0.1, 0.5$ or $0.8$ and distribution truncated at $0.9$ so as to ensure enough signal remains for each variable for the model to work. 

In both the regular and zero-inflated simulations, each configuration is simulated 50 times. 

\subsubsection{Metrics}
 We first want to test the inference procedure's ability to retrieve the model's clustering. We use the adjusted rand index (ARI), computed with package \textbf{aricode} \citep{aricode} to compare ground-truth clustering and inferred clustering. The ARI is a value between $-1$ and $1$ used to compare two clusterings. The higher its value, the closer the clusterings are (an ARI of $1$ meaning they are identical, up to label switching). Its computation is based on the number of elements that are in the same cluster in both cases / in different clusters in both cases / in the same cluster in one case and in two different clusters in the other. For each configuration and each inference method (Normal-block, 2-step method with either residuals-based clustering or variance-based clustering) we compute the median and the standard deviation of the ARI (see Table \ref{tab:ari_res}).

 The inference procedure produces a series of network, one for each value of the penalty $\lambda$ we use for Graphical-Lasso (the higher $\lambda$ gets, the more $0$s $\hat{\Omega}$ contains). In this assessment, we leave aside the issue of chosing $\lambda$. Instead, we compare the real and inferred networks with the Receiving Operator Characteristic (ROC) curve and the corresponding Area Under the Curve (AUC). The ROC curve plots the True Positive rate (or recall) as a function of the False Positive rate (or fallout), AUC is the area under that curve. The larger the AUC, the better is the network reconstruction. Since the model does not change when permuting clusters labels, comparing networks with unobserved clusters requires testing the labels permutations, so that we only compute the AUCs in the observed clusters configuration. 

 To test the model's ability to correctly retrieve other parameters, we use the root mean squared error (RMSE) for $B$, $D$, $\kappa$ (for zero-inflated data). We also use it for $\hat{Y}$ to assess the inference procedure's ability to correctly fit the data.
 
Finally, to assess the computational cost of the various inference methods, we measure
 the execution time required by each one as a function of $n, p$ and $q$.

\subsubsection{Comparison between inference methods}
While we focus on the more elaborate "simultaneous inference" procedure, we compare its results with those obtained with the 2-step approach, using two possible clustering methods (SBM on the precision matrix or k-means on the residuals). When $q = 15$, we do not run the SBM clustering method as it is too computationally expensive.

\subsection{Results}

Table \ref{tab:ari_res} shows that all the inference methods almost systematically retrieve the correct clustering, for all the network structures that we tested. We still note that when $n = 20$, the task becomes harder, especially when $p = 500$ and / or $q = 15$. This indicates that the clustering task gets harder when one has to cluster more entities into more clusters. There is no clear influence of the network structure on the ARI results (see Appendix \ref{appendix_ari_res}). The integrated inference performs slightly better than the 2-step methods in terms of ARI. 

\begin{table}[ht]
    \centering
    \tiny
    \begin{tabular}{|p{0.3cm}|p{0.3cm}|p{0.3cm}|p{1.5cm}|p{1.5cm}|p{1.5cm}|}
        \hline
        n & p & q & Integrated inference - ARI mean (standard deviation) & 2-step method - variance clustering  - ARI mean (standard deviation) & 2-step method - residuals clustering - ARI mean (standard deviation)\\
        \hline
        20 & 100 & 3 & 1 (0.01) & 0.98 (0.08) & 1 (0.01)\\
        20 & 100 & 5 & 0.99 (0.02) & 0.93 (0.10) & 1 (0.01)\\
        20 & 100 & 10 & 0.96 (0.05) & 0.82 (0.13) & 0.97 (0.04)\\
        20 & 100 & 15 & 0.91 (0.07) & NA & 0.91 (0.06)\\
        20 & 500 & 3 & 1 (0) & 0.98 (0.06) & 1 (0)\\
        20 & 500 & 5 & 1 (0) & 0.94 (0.09) & 1 (0)\\
        20 & 500 & 10 & 0.99 (0.01) & 0.86 (0.13) & 0.99 (0.02)\\
        20 & 500 & 15 & 0.97 (0.02) & NA & 0.96 (0.03)\\
        50 & 100 & 3 & 1 (0) & 1 (0) & 1 (0)\\
        50 & 100 & 5 & 1 (0) & 1 (0.02) & 1 (0)\\
        50 & 100 & 10 & 1 (0) & 0.99 (0.03) & 0.98 (0.04)\\
        50 & 100 & 15 & 0.95 (0.16) & NA & 0.93 (0.17)\\
        50 & 500 & 3 & 1 (0) & 1 (0) & 1 (0)\\
        50 & 500 & 5 & 1 (0) & 0.99 (0.04) & 1 (0)\\
        50 & 500 & 10 & 1 (0) & 0.99 (0.02) & 0.99 (0.03)\\
        50 & 500 & 15 & 1 (0) & NA & 0.96 (0.04)\\
        200 & 100 & 3 & 1 (0) & 1 (0) & 1 (0)\\
        200 & 100 & 5 & 1 (0) & 1 (0) & 1 (0)\\
        200 & 100 & 10 & 1 (0) & 1 (0) & 0.99 (0.03)\\
        200 & 100 & 15 & 0.98 (0.12) & NA & 0.95 (0.10)\\
        200 & 500 & 3 & 1 (0) & 1 (0) & 1 (0)\\
        200 & 500 & 5 & 1 (0) & 1 (0) & 1 (0)\\
        200 & 500 & 10 & 1 (0) & 1 (0) & 0.97 (0.05)\\
        200 & 500 & 15 & 1 (0) & NA & 0.95 (0.03)\\
        500 & 100 & 3 & 1 (0) & 1 (0) & 1 (0)\\
        500 & 100 & 5 & 1 (0) & 1 (0) & 1 (0)\\
        500 & 100 & 10 & 1 (0) & 1 (0) & 0.99 (0.03)\\
        500 & 100 & 15 & 1 (0) & NA & 0.95 (0.03)\\
        500 & 500 & 3 & 1 (0) & 1 (0) & 1 (0)\\
        500 & 500 & 5 & 1 (0) & 1 (0) & 1 (0)\\
        500 & 500 & 10 & 1 (0) & 1 (0) & 0.97 (0.05)\\
        500 & 500 & 15 & 1 (0) & NA & 0.94 (0.04)\\
        \hline
    \end{tabular}
    \caption{ARI results for each configuration: its median and standard deviation are shown for each method.The table only shows the results for the "preferential attachment" network structure. Results for the other two network structures (Erdös-Rényi and Community) are given in appendix \ref{appendix_ari_res}.}
    \label{tab:ari_res}
\end{table}

\begin{figure}[h!]
    \centering
    \includegraphics[ height = 7cm]{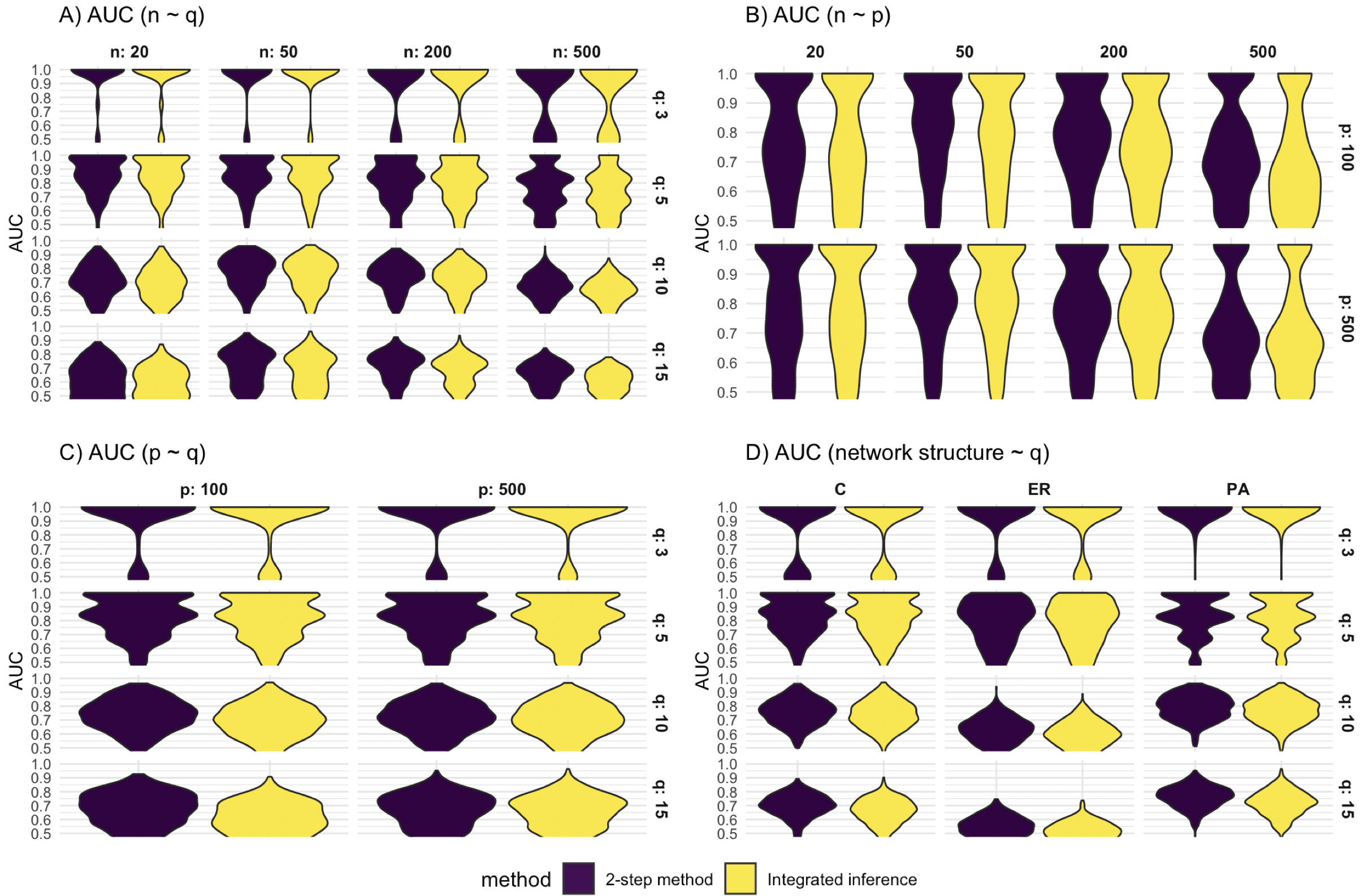}
    \caption{Violin plots of the AUC for non-zero-inflated data, for different network structures and different values of $n$, $p$ and $q$. For the network structure, \textbf{C} stand for community, \textbf{ER} for Erdös-Rényi, and \textbf{PA} for preferential attachment.}
    \label{fig:new_AUC_violin_no_zi}
\end{figure}

Figure \ref{fig:new_AUC_violin_no_zi} shows that both the integrated and the 2-step method correctly retrieve the network structure. However, we see that as the number of clusters $q$ increases, the graph structure is harder to retrieve. This is likely owing to the fact that the network size increases whereas the number of variables per cluster does not. We also see that the Erdös-Rényi networks (no particular structure) are harder to retrieve than the more structured ones when $q = 10$.

\begin{figure}[h!]
    \centering
    \includegraphics[height = 6cm]{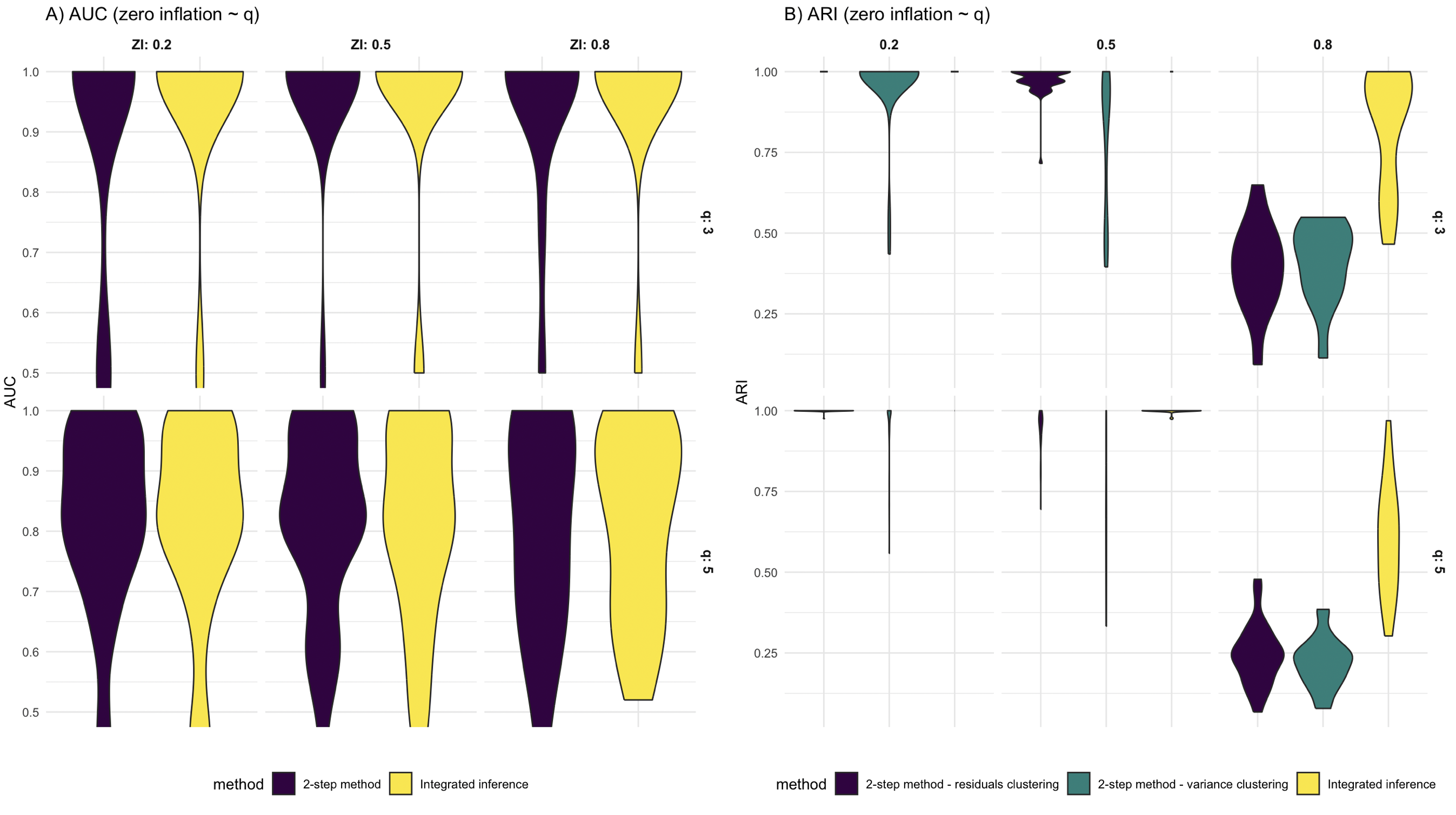}
    \caption{Violin plots of the AUC (A) (observed clustering) and ARI (B) for zero-inflated data, for different zero-inflation levels and values of $q$.}
    \label{fig:new_ZI_AUC_ARI}
\end{figure}

In the case of zero-inflated data, Figure \ref{fig:new_ZI_AUC_ARI} A shows that the AUC is only slightly worse when the zero-inflation increases. However, a more important zero-inflation significantly affects the model's ability to retrieve the correct clustering ( Figure \ref{fig:new_ZI_AUC_ARI} B). In terms of ARI, we also see that the 2-step methods' results are much worse than those of the integrated inference method when the zero-inflation is important.

\begin{figure}[h!]
    \centering
    \includegraphics[height = 6cm]{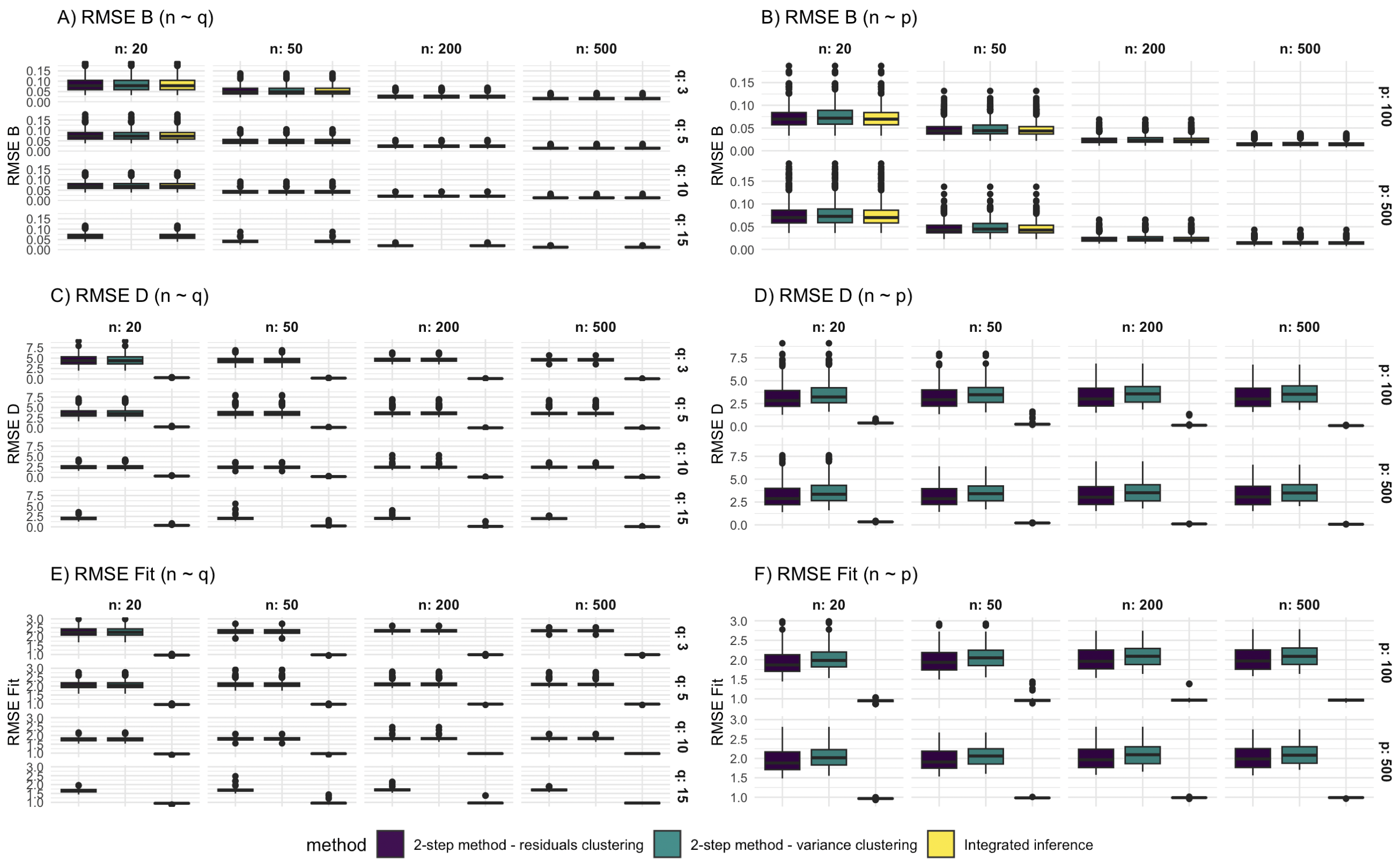}
    \caption{RMSE measured for $B$ (A, B), $D$ (C, D) and for real-data fitting (E, F) for different values of $n$, $p$ and $q$ (non-zero-inflated data). On each plot, the first column corresponds to the 2-step method with residuals clustering, the second to the 2-step method with variance clustering and the third to the integrated inference method.}
    \label{fig:new_RMSE_boxplots_no_zi}
\end{figure}

\begin{figure}[h!]
    \centering
    \includegraphics[height = 8cm]{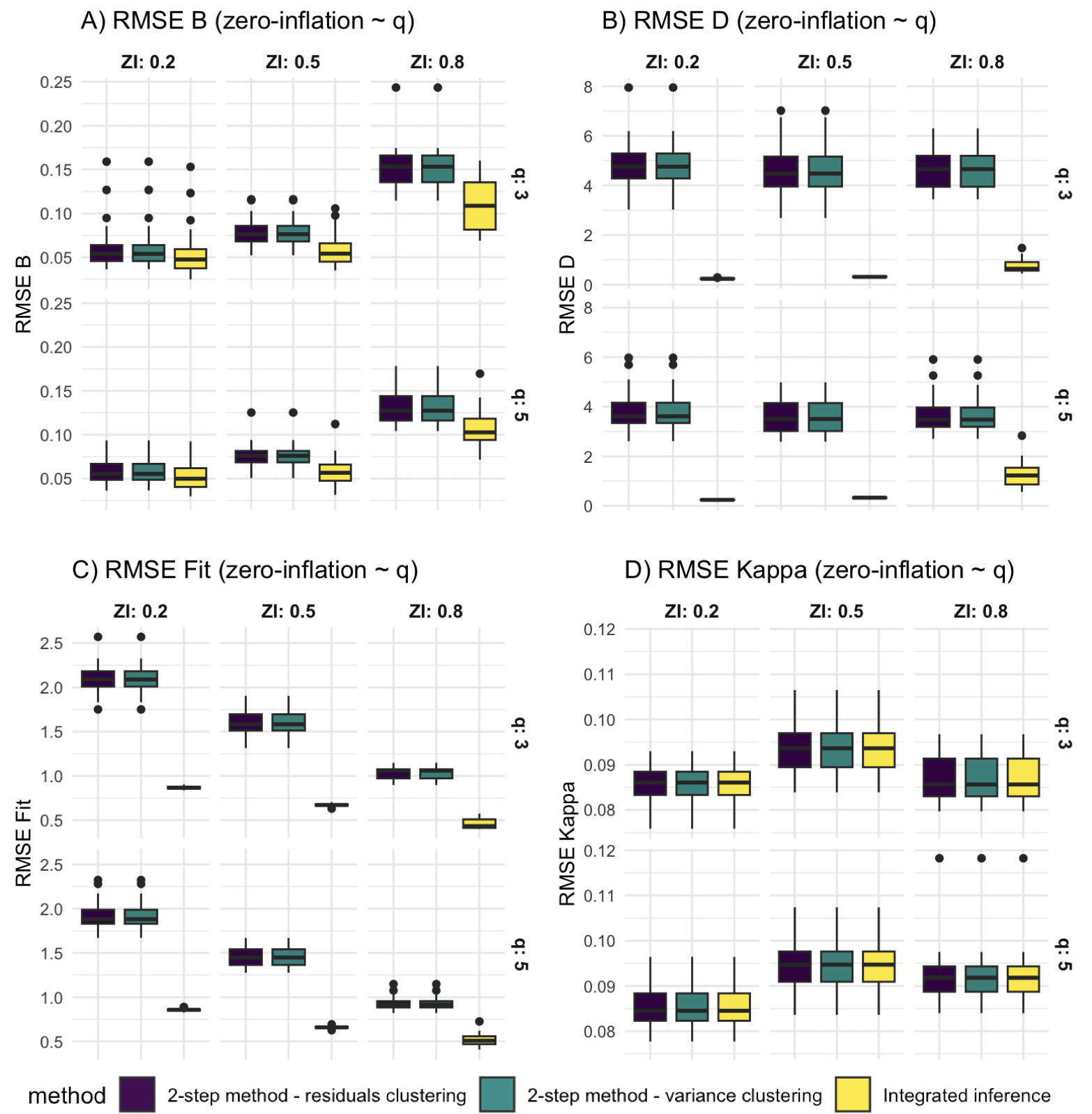}
    \caption{RMSE measured for zero-inflated data, for $B$ (A), $D$ (B), for real-data fitting (C)  and for $\kappa$ (D) for different values of $q$ and different zero-inflation levels.}
    \label{fig:new_ZI_RMSE}
\end{figure}

Figure \ref{fig:new_RMSE_boxplots_no_zi} hows that increasing $n$ quite logically reduces the error on the estimates of $B$ for all inference methods, whereas increasing $p$ or $q$ does not seem to have a significant effect. $D$ is always better estimated with the integrated inference method, which makes sense as the 2-step methods are not designed to tease apart its effect on the observations. The data is also better fitted with the integrated inference method.

In the case of zero-inflated data (Figure \ref{fig:new_ZI_RMSE}), an increasing zero-inflation increases the error for $B$ while it reduces it for the fitting error. This might be explained by the increasing number of zeros that are correctly predicted.  For $D$, the 2-step methods error does not seem significantly impacted by an increased zero-inflation whereas the RMSE increases for the integrated inference.

\begin{figure}[ht]
    \centering
    \includegraphics[height = 8cm]{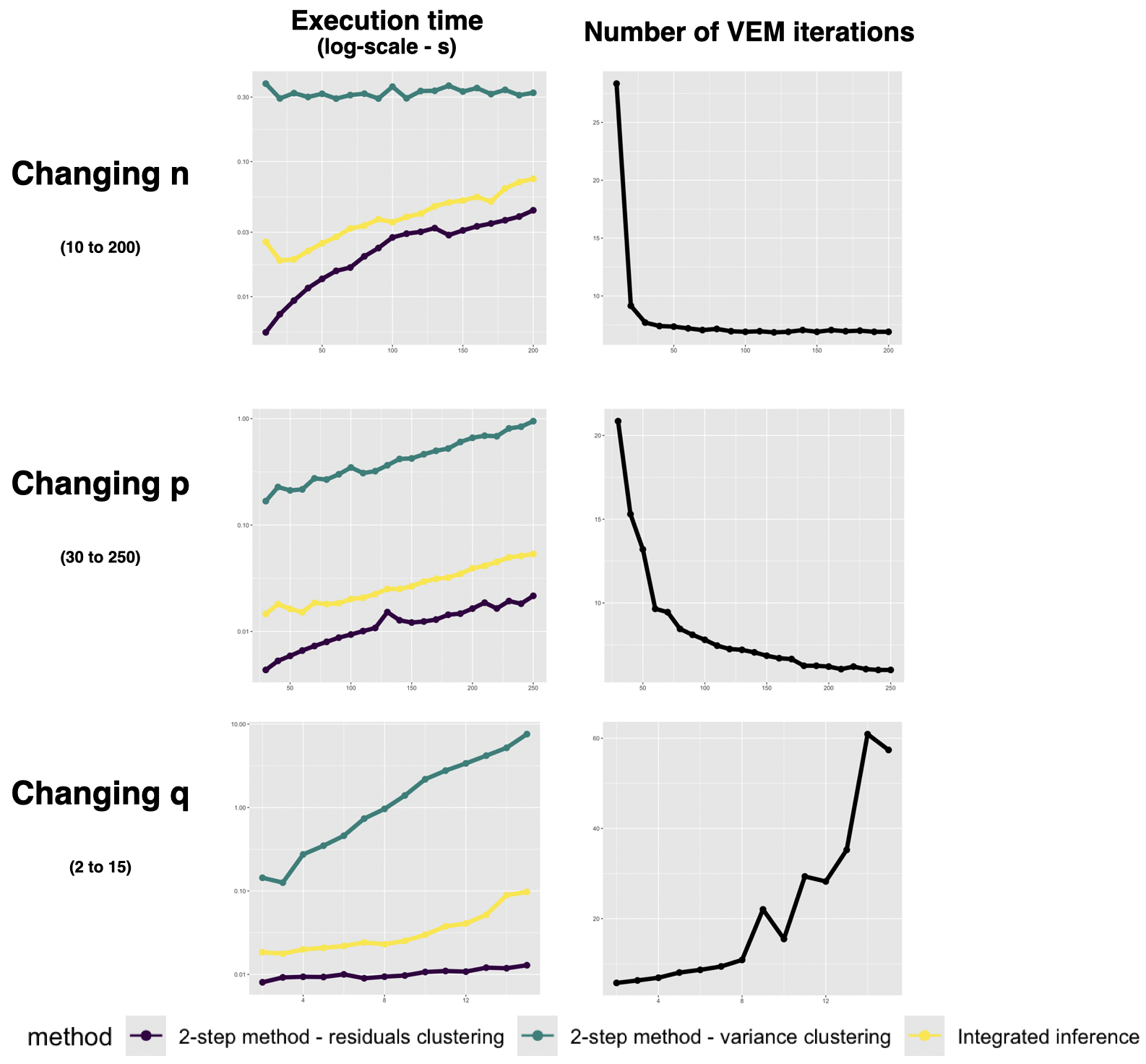}
    \caption{Execution time (log-scale) and number of VEM iterations (for the integrated inference approach) as a function of $n, p$ and $q$ (when one varies, the others are fixed to $n = 30$, $p = 100$, $q = 5$). Each configuration is simulated $20$ times and we plot the median result over these simulations. }
    \label{fig:execution_time}
\end{figure}

Figure \ref{fig:execution_time} shows the execution time required for different configurations. We see that the 2-step inference procedure with a variance-based clustering is systematically the one that takes longer. This is probably owing to the computational cost of the Stochastic Block Models algorithm. The integrated inference approach necessarily takes longer than the 2-step methods with residuals clustering since the latter is used to initialize the former. Interestingly, on the one hand, increasing $n$ or $p$ reduces the number of VEM iterations required for convergence while not reducing the execution time. It can be explained by the fact that increasing $n$ or $p$ makes the clustering task easier since more information is given but each matrix operation is more costly. On the other hand, increasing $q$ induces an increase in the number of iterations and in the execution time.

\begin{figure}[h!]
    \centering
    \includegraphics[height = 6cm]{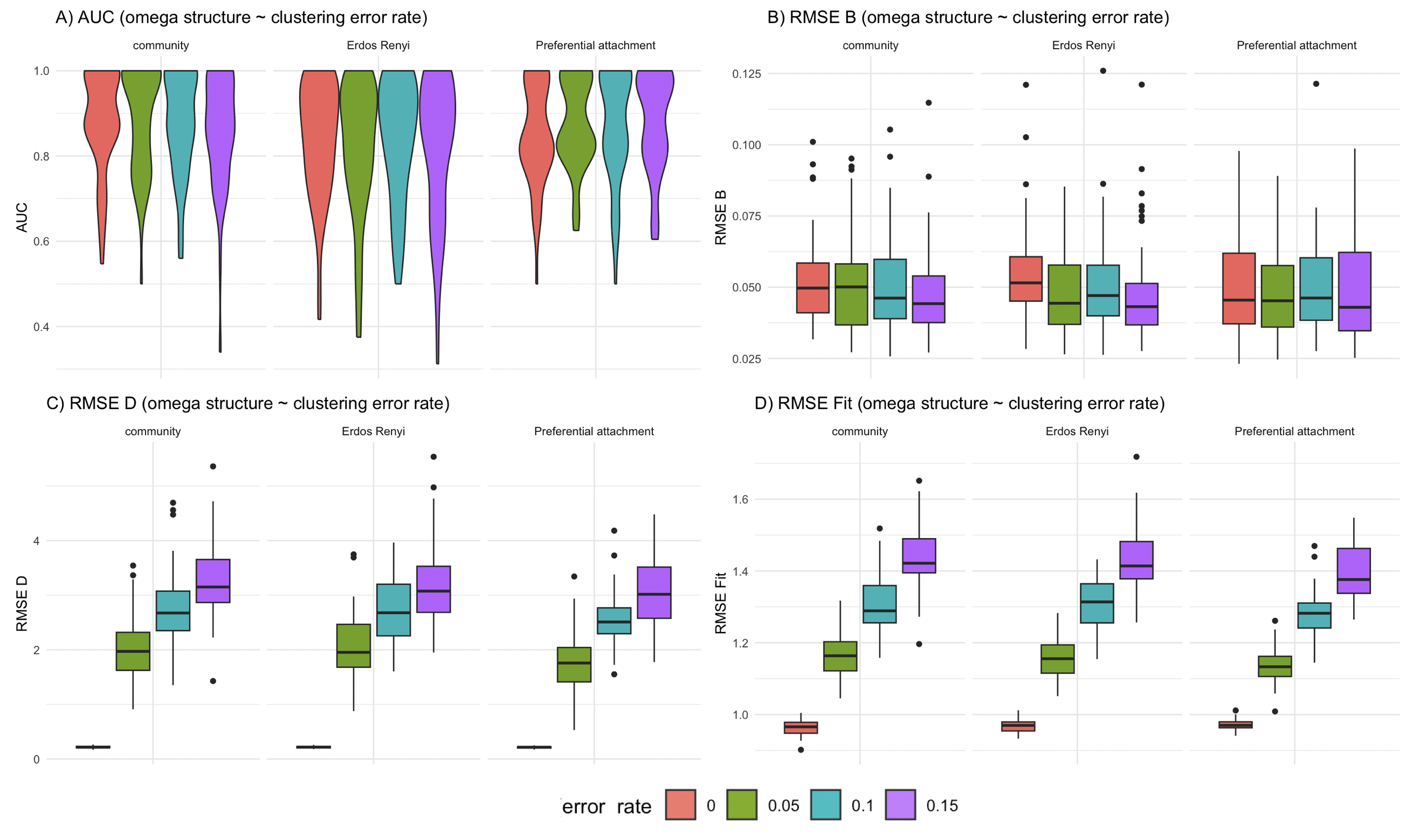}
    \caption{AUC and RMSE results when a wrong clustering is given in the inference procedure. The error rate corresponds to the fraction of variables for which a wrong clustering label is given.}
    \label{fig:new_wrong_clustering_results}
\end{figure}

Regarding the simulations with an erroneous clustering, Figure \ref{fig:new_wrong_clustering_results} shows that the AUC and the RMSE for the regression matrix $B$ are not much impacted by the clustering mistakes. However, an increase in the error rate leads to significantly worse results in terms of RMSE for $D$ and for the data fitting.

\newpage
\section{Illustrations}\label{illustration}

We show how the different variants of the models can be applied to analyse data from different fields. First we use a simple Normal-Block model for the analysis of proteomics data and show how the clustering it outputs may help retrieve connections between different pathways. We also use a regularized Normal-Block model to analyse words occurrences on webpages and how the different groups of words tend to be found together or not. Data and code for these illustrations are available in the \textit{inst} folder
of the \textbf{normalblockr} github repository.

\subsection{Breast cancer proteomics data}
\begin{figure}[ht]
    \centering
    \includegraphics[width = 10cm]{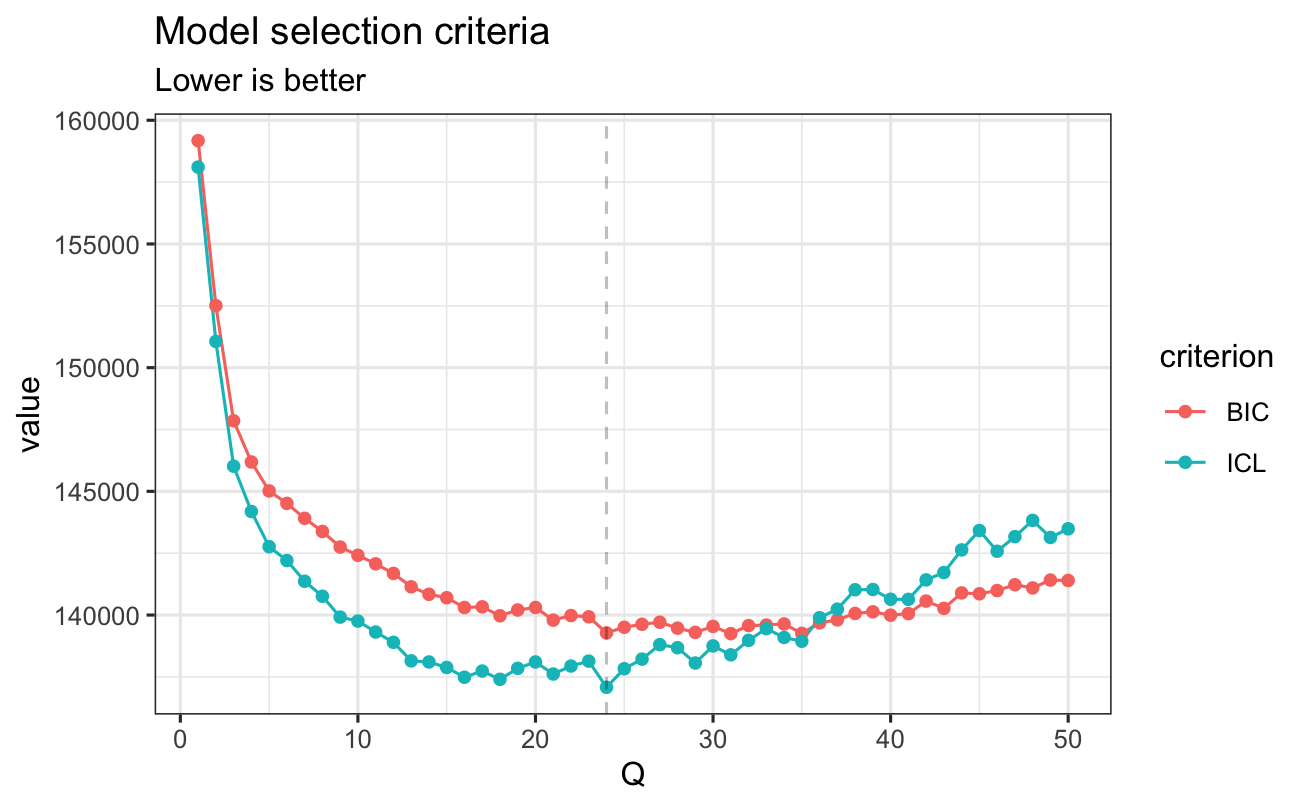}
    \caption{BIC and ICL criteria as a function of the number of clusters $q$ for the integrated inference method applied to the breast cancer dataset.}
    \label{fig:tcga-criteria}
\end{figure}

We first use Normal-Block to analyse proteomics data of breast cancer from Brigham and Women's Hospital [2012] obtained with reverse-phase protein arrays. We pre-process the data to remove proteins whose expressions are highly correlated. We also remove the sites (that is, tumors here) that appear as outliers on a PCA of the proteins expressions. We consider the standardized expressions of $p = 163$ proteins in $n = 346$ tumors. We use the breast cancer subtypes as covariates (Normal-Like, Basal, Luminal A, Luminal B and HER2-enriched) and run the integrated inference from $q = 1 $ to $q = 50$ clusters. Using the ICL, we fix $q = 24$ clusters (Figure \ref{fig:tcga-criteria}). 

\begin{figure}[ht]
    \centering
    \includegraphics[width = 15cm]{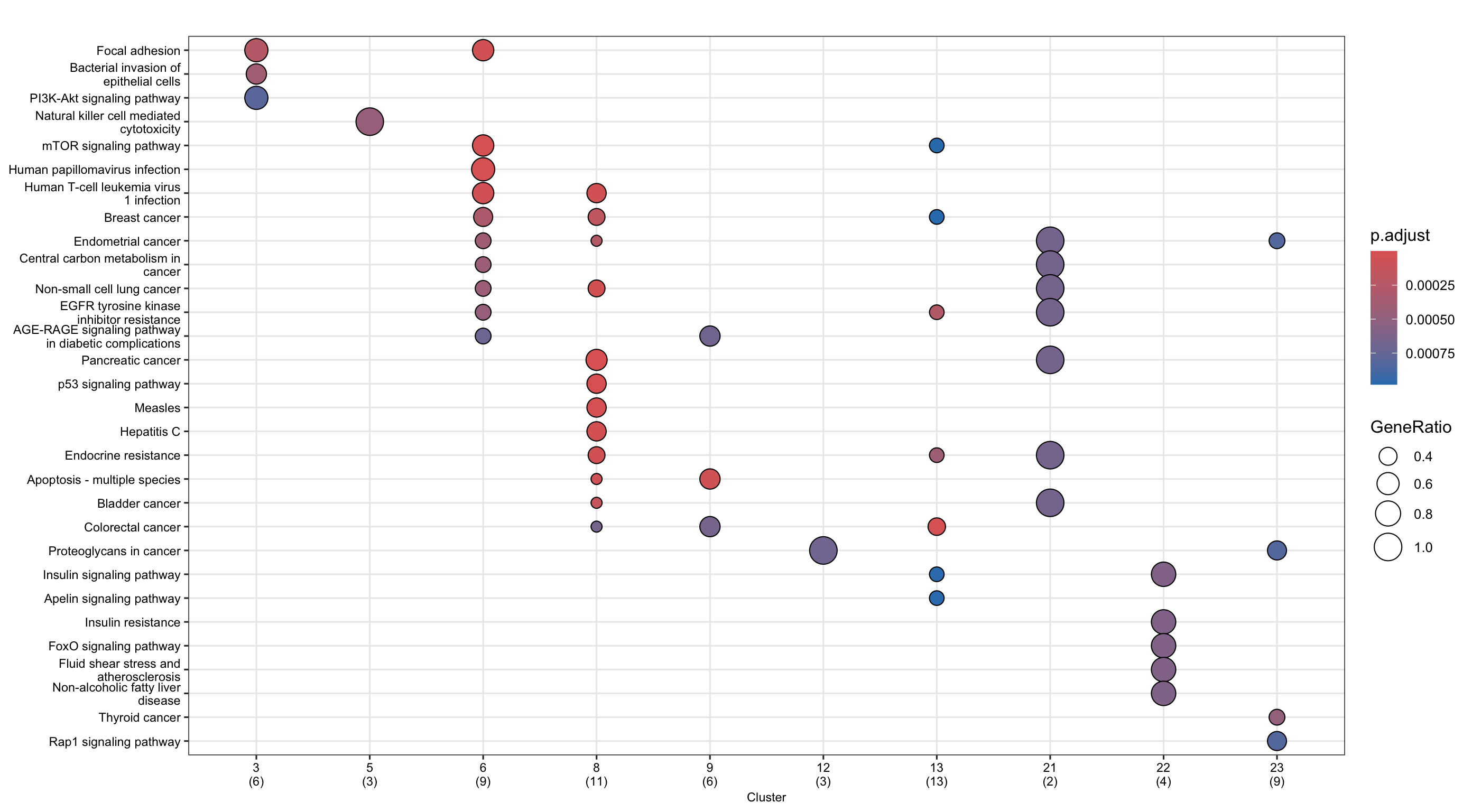}
    \caption{Enrichment plot for the clusters obtained with Normal-Block. Below the cluster labels is given the number of genes from the cluster that is associated with one of the identified pathways.}
    \label{fig:tcga-enrichment}
\end{figure}

We then run an enrichment analysis based on the Kyoto Encyclopedia of Genes and Genomes (KEGG) on the resulting clusters with a p-value threshold of $01$, using R packages \textbf{clusterProfiler} and \textbf{enrichplot} \citep{yu2012clusterprofiler, yu2021enrichplot}. The results are shown on Figure \ref{fig:tcga-enrichment}. 10 clusters out of 24 display distinct enrichment patterns that pass the p-value threshold. Some of the pathway combinations appear meaningful from a biological point of view. For instance, PI3K-Akt and focal adhesion pathways are identified together in cluster 3 and they happen to be related in certain cancer cells \citep{matsuoka2012pi3k}. Another example is the identification of both EGFR and mTOR pathways in clusters 6 and 13, in breast cancers, the overexpression of EGFR is associated with the activation of the PI3K/Akt/mTOR pathway \citep{matsuoka2012pi3k}

\subsection{University webpages}
 Following \cite{tan2015cluster}'s illustration for their cluster Graphical-Lasso, we use Normal-Block for words frequencies analysis on webpages from the "4 universities data set" accessible on \textbf{www.cs.cmu.edu}. This dataset from the “World Wide Knowledge Base” project at Carnegie Mellon University gathers web pages from four universities: Cornell, Texas, Washington, and Wisconsin. We process the data as indicated in \cite{tan2015cluster} but they used pre-processed data \cite{cardoso_cachopo} that we could not have access to. 

 We consider students webpages only, leaving us with $n = 504$ pages and $p = 1867$ words after removal of stop words and of words occurring only once in the whole dataset. 

\begin{figure}[h!]
    \centering
    \includegraphics[width = 6cm]{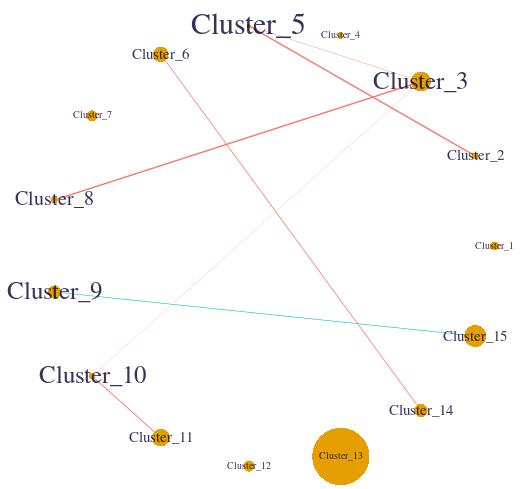}
    \caption{Network obtained with Normal-Block for the "4 universities dataset". Pink edges correspond to positive associations whereas blue edges are for negative associations, the thicker an edge, the stronger the corresponding association. }
    \label{fig:4uni-network}
\end{figure}

 Let $f_{ij}$ be the frequency of the $j-th$ word on the $i-th$ page. We consider $\tilde{Y} \in \mathbb{R}^{n \times p}$ defined by $Y_{ij} = \log(1 + f_{ij})$. We only keep the $p = 100$ words with maximal entropy, with entropy for the $j$-th term defined as $- \sum_{i = 1}^n g_{ij} \log(g_{ij}) / \log(n)$ with $g_{ij} = \frac{f_{ij}}{\sum{i=1}^n f_{ij}}$. We then obtain $Y$ from $\tilde{Y}$ standardizing each column to have mean $0$ and standard deviation $1$. We run the model with $q = 15$ clusters and a sparsity penalty $\lambda = 05$. The clustering is described in table \ref{table:4uni-clustering} and the network is shown on Figure \ref{fig:4uni-network}.

\begin{table}[ht]\label{table:4uni-clustering}
\centering
\begin{tabular}{cl}
  \hline
 cluster & words \\ 
  \hline
 1 & research, thu, data, performance \\ 
 2 & postscript, available, game \\ 
 3 & systems, system, proceedings, distributed, report, operating, conference, \\ &  workshop, paper \\ 
 4 & seattle, class, summer \\ 
 5 & work, project \\ 
 6 & also, software, like, web, stuff, date, david \\ 
 7 & madison, wisconsin, sep, usa, really \\ 
 8 & appear, international, james \\ 
 9 & home, page, office, phone, email, number \\ 
 10 & parallel, programming, group \\ 
 11 & interests, theory, one, compiler, language, think, design, languages \\ 
 12 & department, time, cornell, homepage, seed \\ 
 13 & nov, can, monday, student, graduate, engineering, working, wednesday, links, \\ & oct, jan, currently, may, current, new, school, fall, will, see, \\ & java, first, year, interesting, server, database, algorithms \\ 
 14 & people, computers, two, well, make, program \\ 
 15 & computer, university, austin, science, information, texas, address, sciences, \\ & using, learning \\ 
   \hline
\end{tabular}
\caption{List of Words by cluster in the "4 universities dataset"} 
\end{table}

 As in \cite{tan2015cluster}, we see that words like "office", "phone" and "email" or "student" and "graduate" tend to be grouped together so that our clustering seems consistent with the one they obtain. Other groupings are meaningful such as that of "conference", "workshop" and "paper" in cluster 3, "parallel" and "programming" in cluster 10 or "austin" and "texas" in cluster 15 . 

 In the network we see on the one hand that the group of generic "administrative" words represented by cluster 9 tend not to be found with other more computer-science related words in cluster 15. On the other hand, clusters 8 and 3 display a positive association, maybe because they both relate to scientific communication.

\newpage
\section{Discussion}

We propose Normal-Block, a Gaussian graphical model that integrates a clustering on the variables. This adds structure to the model. Moreover, considering a network at cluster level reduces the network's dimensionality. We prove the model's identifiability when the clustering is observed and propose an inference procedure that resorts to Graphical-Lasso and uses variational expectation-maximization to simultaneously retrieve the clustering and the cluster-level-network. 

A limitation of Graphical-Lasso and other penalized methods for network inference lies in the lack of structure they assume, as well as the complexity associated with network inference and interpretation as its dimension grows. The Normal-Block adds a structure hypothesis on the network and builds on this hypothesis to reduce the network's dimension. 

We have shown on simulated data that both the clustering and the network inference work well, provided the zero-inflation remains limited, but we have no theoretical results on the model's identifiability when the clustering is not observed or on the global concavity of the ELBO. 

The model could further be developed to introduce other forms of zero-inflation, for instance making it individual-dependent or covariate-dependent instead of variable-dependent. It could also be useful to add other \textit{a priori} hypotheses that would constrain the network structure, for instance forbidding or favouring specific associations to appear in the network. Finally, several temporal extensions of the model could be designed, to consider a clustering or a network that could evolve at each time step.

\section*{Acknowledgements}
We thank Mahendra Mariadassou for his insights in the analysis of the model's identifiability.
This work was partially supported by a funding of the ANR SingleStatOmics.

\newpage
\nocite{*}

\appendix
\section*{Appendices}
\renewcommand{\thesection}{\Alph{section}}

\section{Proofs of identifiability }\label{appendix_identifiability}
\subsection{Observed clusters model}
\subsubsection{Spherical model}

In this model we have $Y_i ~ p_{\theta}(Y_i) = \mathcal{N}(\mu_i = B^{\top}X_i, S = D + C \Sigma C^{\top})$ with $D = \mathrm{diag}(\xi), \xi \in \mathbb{R}^+$ and the model parameters $\theta = (\xi,B,\Sigma)$. Let us assume that no group is empty.

 Given two sets of parameters $\theta, \theta'$, for $i \in [\![ 1 ; n ]\!]$:
\begin{align*}
    & P_{\theta}(Y_i) = P_{\theta'}(Y_i) \\
    \Leftrightarrow  & \| Y_i - \mu\|_{S^{-1}} - \log |S| = \| Y_i - \mu'\|_{S'^{-1}} - \log |S'| \\
    \Leftrightarrow & Y_i^T(S^{-1} - S'^{-1})Y_i - 2Y_i^T (S^{-1}\mu  -  S'^{-1}\mu') \\
    & \quad \qquad + \mu^T S^{-1} \mu - \mu'^T S'^{-1} \mu' + \log |S| - \log |S'| = 0
\end{align*}
which is equal to zero when $\xi I_p + C \Sigma C^T = \xi' I_p  + C \Sigma' C^T$ and $\mu_i = \mu_i' \Leftrightarrow (B - B')^{\top} X_i = 0$. Having $\forall i \in [\![ 1 ; n ]\!], (B - B')^{\top} X_i = 0$ is equivalent to $(B - B')^{\top} X = 0$. Thus, if $X$ is of rank $d$, we have $B = B'$.

 For the variance, let us abusively denote $C(j) \in [\![1; q]\!]$ the cluster $j$ belongs to. We assume that no group is empty so that:
\begin{align*}
    \forall k_1^*, k_2^* \in [\![1 ; q]\!], k_1^* \neq k_2^*, & \exists j, l \in [\![1 ; p]\!], C(j) = k_1^*, C(l) = k_2^* \\
    (C \Sigma' C^{\top})_{jl}) &=  \sigma'_{k_1^*k_2^*} = \sigma_{k_1^* k_2^*}
\end{align*}
 This proves the equality between $\Sigma$ and $\Sigma'$ off-diagonal terms. 

 On the diagonal of $S'$:
\begin{align*}
    \forall j \in [\![1 ; p]\!], S'_{jj} &= \xi' + \sigma'_{C(j) C(j)} = \xi + \sigma_{C(j) C(j)}
\end{align*}

 Let us  assume that there exists one group $k^*$ that contains at least two elements.
\begin{align*}
    \exists j^*, l^* \in [\![1 ; p]\!], & j^* \neq l^*, C(j^*) = C(l^*) = k^* \\
    (C \Sigma' C^{\top})_{j^*l^*} &=  \sigma'_{k^* k^*} = \sigma_{k^* k^*} \\
    S'_{j^*j^*} & = \xi' + \sigma'_{k^* k^*} \\
    &= \xi' + \sigma_{k^* k^*}
\end{align*}

 Thus we have $\xi = \xi'$. 

 The diagonal expression $\xi' + \sigma'_{C(j) C(j)} = \xi + \sigma_{C(j) C(j)}$ finally gives us the equality between $\Sigma$ and $\Sigma'$ diagonal terms. 

 Therefore, the model is identifiable as long as $X$ is a full-rank matrix, no cluster is empty and at least one of the clusters contains at least two elements.

\subsubsection{Diagonal model}

In this model we have $Y_i \sim p_{\theta}(Y_i) = \mathcal{N}(\mu = B^{\top}X_i, S = D + C \Sigma C^{\top})$ with $D = \mathrm{diag}(d), d \in \mathbb{R}^{+^p}$ and the model parameters $\theta = (d,B,\Sigma)$. Let us assume that no group is empty.

 Given two sets of parameters $\theta, \theta'$:
\begin{align*}
    & P_{\theta}(Y_i) = P_{\theta'}(Y_i) \\
    \Leftrightarrow  & \| Y_i - \mu\|_{S^{-1}} - \log |S| = \| Y_i - \mu'\|_{S'^{-1}} - \log |S'| \\
    \Leftrightarrow & Y_i^T(S^{-1} - S'^{-1})Y_i - 2Y_i^T (S^{-1}\mu  -  S'^{-1}\mu') \\
    & \quad \qquad + \mu^T S^{-1} \mu - \mu'^T S'^{-1} \mu' + \log |S| - \log |S'| = 0
\end{align*}
which is equal to zero when $\xi I_p + C \Sigma C^T = \xi' I_p  + C \Sigma' C^T$ and $\mu = \mu' \Leftrightarrow B^T X_i = B'^T X_i \Rightarrow B = B'$ and $X_i^T X_i$ non singular (thus full-rank $X_i$). 

 For the variance, we assume that no group is empty so that:
\begin{align*}
    \forall k_1^*, k_2^* \in [\![1 ; q]\!], k_1^* \neq k_2^*, & \exists j, l \in [\![1 ; p]\!], C(j) = k_1^*, C(l) = k_2^* \\
    (C \Sigma' C^{\top})_{jl}) &=  \sigma'_{k_1^* k_2^*} = \sigma_{k_1^* k_2^*}
\end{align*}
This proves the equality between $\Sigma$ and $\Sigma'$ off-diagonal terms.

 Let us also assume that each group $q$ contains at least two elements.
\begin{align*}
    \forall k^* \in [\![1 ; q]\!], & \exists j, l \in [\![1 ; p]\!], j \neq l, q(j) = q(l) = k^* \\
    (C \Sigma' C^{\top})_{jl} &=  \sigma'_{k^* k^*} = \sigma_{k^* k^*}
\end{align*}
 This proves the equality between $\Sigma$ and $\Sigma'$ diagonal terms, provided each group contains at least two elements. 

 Finally on the diagonal of $S'$':
\begin{align*}
    \forall j \in [![1 ; p]\!], S_{jj} &= d_j + \sigma'_{q(j) q(j)} \\
    &= d_j + \sigma_{q(j) q(j)} \\
    &= d_j +  \sigma_{q(j) q(j)}
\end{align*}
 This proves that $d_j' = d_j$. Thus, the model is identifiable provided each group contains at least two elements. 

 If one group $k^*$ contains only one element $j$ however, any $d_j+ \epsilon, \epsilon > - d_j, \epsilon < \sigma_{k^* k^*}$ can give the same likelihood, replacing $\sigma_{k^* k^*}$ with $\sigma'_{k^* k^*} = \sigma_{k^* k^*} - \epsilon$.

 Therefore the observed-clusters model is identifiable provided the $X_i$ are full rank and each cluster contains at least two elements. 

\subsection{Unobserved clusters model}
The model's parameters are $\theta = (B, \Sigma, D, \alpha)$. We want to prove that if $\forall i \in [\![1 ; n]\!], p_{\theta}(Y_i) = p_{\theta'}(Y_i)$ then $\theta = \theta'$.Let us denote $\sigma_k$ the $k$-th diagonal element of $\Sigma$ and $\sigma_{k_1 k_2} := \Sigma_{q_{k_1 k_2}}$. Finally, if $s$ is a permutation of $[\![1 ; q]\!]$, we define $\Sigma^{(s)}$ by $\Sigma^{(s)}_{q_{jl}} = \Sigma_{q_{s(j)s(l)}}$ and $\alpha^{(s)}$ by $\alpha^{(s)}_j = \alpha_{s(j)}$ We make the following mild hypotheses about the parameters:

\begin{enumerate}
    \item $p > q$ \textbf{(H1)} 
    \item $X$ is a full-rank matrix \textbf{(H2)}.
    \item $\forall k \in [\![1 ; q]\!], \alpha_k > 0$  \textbf{(H3)}.
    \item $\forall k_1, k_2 \in [\![1 ; q]\!], k_1 \neq k_2 \Rightarrow \sigma_{k_1} \neq \sigma_{k_2}$ (that is to say the diagonal values of $\Sigma$ are distinct two by two)  \textbf{(H4)}.
    \item $\forall k_1, k_2, k_3 \in [\![1 ; q]\!], k_2 \neq k_3 \Rightarrow \sigma_{k_1} \neq \sigma_{k_2 k_3}$ (that it to say there is non non-diagonal value of $\Sigma$ that is equal to one of its diagonal values)  \textbf{(H5)}. 
\end{enumerate}

For $i \in [\![1 ; n]\!]$, we have:

\begin{align*}
    p_{\theta}(Y_i) = \sum_{C^* \in [\![1; q ]\!]^p} \left(\Pi_{j=1}^p \alpha_{C_j^*}\right) \mathcal{N}(Y_i | B^{\top}X_i, D + C^* \Sigma_{q} C^{* \top}),
\end{align*}

Let us arbitrarily sort the $C^* \in [\![1; q ]\!]^p$, from $a = 1$ to $a = q^p$ and denote them $(C_a)_{1 \leq a \leq q^p}$. Let us abusively denote $C_{a}(j)$ the cluster of $j \in [\![1; p]\!]$ in the clustering defined by matrix $C_a$. Then $p_{\theta}(Y_i)$ can be rewritten:

\begin{equation}
\label{eq:mixture-coefficients}    
\begin{aligned}
    p_{\theta}(Y_i) &= \sum_{a = 1}^{q^p} \left(\Pi_{j = 1}^p \alpha_{C_a(j)}\right) f(Y_i; B^{\top}X_i, D + C_a \Sigma C_a^{\top}) \\
    &= \sum_{a = 1}^{q^p} \gamma_a f(Y_i ; \mu_i, \Sigma_a),
\end{aligned}
\end{equation}

where $f$ denotes the probability distribution function of a multivariate Gaussian distribution and the mixture parameteres are given by $\gamma_a = \Pi_{j = 1}^p \alpha_{C_a(j)}$, $\mu_i = B^{\top}X_i$ and $\Sigma_a = D + C_a \Sigma C_a^{\top}$. \\

Hypothesis $(3)$ guarantees that no two $\Sigma_a$ can be equal as two distinct values of $a$ correspond to two different clustering and at least one diagonal value of $\Sigma_a$ is consequently modified. Thus, in the rewritten expression of the likelihood $(1)$, one recognizes a finite mixture of Gaussian distribution with two by two distinct sets of parameters. \cite{yakowitz1968identifiability} proved the identifiability of such a mixture model. Thus, if there exists $(\mu_i, (\gamma_a, \Sigma_a)_{1 \leq a \leq q^p}) $ and $(\mu_i', (\gamma^{'}_{a'}, \Sigma^{'}_{a'})_{1 \leq a' \leq q^p}) $ such that $\sum_{a = 1}^{q^p} \gamma_a f(Y_i ; \mu_i, \Sigma_a)= \sum_{a' = 1}^{q^p} \gamma_{a'} f(Y_i ; \mu_i', \Sigma^{'}_{a'}) $ then:

\begin{align}
\label{eq:mixture-indentifiability}
\begin{split}
    & \forall a \in [\![1; q^p ]\!], \exists ! a'  \in [\![1; q^p ]\!], \gamma_a = \gamma^{'}_{a'}, \Sigma_a = \Sigma^{'}_{a'} \\
    & \forall i \in [\![1; n ]\!], \mu_i = \mu_i'
\end{split}
\end{align}

Let us then assume that there exists $\theta, \theta'$ such that $\forall i \in [\![1 ; n]\!], p_{\theta}(Y_i) = p_{\theta'}(Y_i)$.  $((\mu_i)_{1 \leq i \leq n}, (\gamma_a, \Sigma_a)_{1 \leq a \leq q^p}))$ and  $(\mu_i', (\gamma^{'}_{a'}, \Sigma^{'}_{a'})_{1 \leq a' \leq q^p}) $ denote the parameters that correspond to Eq.~\eqref{eq:mixture-coefficients}. \\

This first implies that $\forall i  \in [\![1; n ]\!], \mu_i = \mu_i'$. Since $X$ is a full-rank matrix  \textbf{(H2)}, this implies that $B = B'$, just as in the observed clusters situation.\\

Up to reordering of the terms in $(\mu_i', (\gamma^{'}_{a'}, \Sigma^{'}_{a'})_{1 \leq a' \leq q^p})$ and using a different clustering sorting in $\theta$ and $\theta'$, we can assume without loss of generality that $\gamma_a = \gamma^{'}_{a}$ and  $\Sigma_a = \Sigma^{'}_{a}$. Note $C_a$ (resp. $C'_a$) the clustering corresponding to $(\Sigma_a, \gamma_a)$ in $\theta$ (resp. in $\theta'$).\\

Now let us prove that there exists a single permutation of $[\![1 ; q]\!]$, denoted $s$ that maps the clusters from $\theta$ to those of $\theta'$, or formally such that $\forall a \in [\![1; q^p, ]\!] \forall j \in [\![1; p ]\!], C^{'}_{a}(j) = s(C_a(j))$. \\
Let us consider $j \in [\![1 ; p]\!]$ and  $a,b \in [\![1 ; q^p]\!]$ and prove that $C_a(j) = C_b(j) \Leftrightarrow C'_a(j) = C'_b(j)$. Consider first $a,b$ such that $C_a(j) = C_b(j)$. Looking at the $j$-th diagonal terms of $\Sigma_a$ and $\Sigma_b$, we have $d'_j + \sigma'_{C'_a(j)} = \Sigma_{a_{jj}} = d_j + \sigma_{C_a(j)} = d_j + \sigma_{C_a(j)} = \Sigma_{b_{jj}} = d'_j + \sigma'_{C'_b(j)}$ and therefore $\sigma'_{C^{'}_{a}(j)} = \sigma'_{C^{'}_{b}(j)}$. Thanks to hypothesis \textbf{(H4)}, this means that $C^{'}_{a}(j) = C^{'}_{b}(j)$. Likewise, using the same arguments, if $a,b$ are such that $C'_a(j) = C'_b(j)$ then $C_a(j) = C_b(j)$.\\
Since $C_a(j)$ reaches all the values of $[\![1 ; q]\!]$ , this proves the existence of a permutation $s_j$ of $[\![1 ; q]\!]$ such that $\forall a \in [\![1 ; q^p]\!], C^{'}_{a}(j) = s_j(C_{a}(j))$. \\

Let's now prove prove that $s_j = s_1 := s$ for all $l \in [\![1 ; p]\!]$. Assume the existence of a $j$ such that $s_j \neq s_1$. Consider $q_0$ any cluster such that $s_j(q_0) \neq s_1(q_0)$, consider the clustering $a$ defined by $C_{a}(.) = q_0$. Since $p > q$, \textbf{(H1)}, by the pigeonhole principle there exist two indexes $k \neq l$ (one of them potentially equal to $1$ and $j$) such that $s_k(q_0) = s_l(q_0) = q_1$. Then by definition of $a$ and $s_k, s_l$, $\Sigma_{a_{1j}} = \sigma_{q_0} = \Sigma_{{a}_{kl}}$. But using the relation $\Sigma_{a_{kl}} = \Sigma'_{a_{kl}}$, we also have $\sigma'_{s_1(q_0)s_j(q_0)} = \Sigma_{a_{1j}} = \Sigma_{{a}_{kl}} = \sigma'_{s_k(q_0)s_l(q_0)} = \sigma'_{q_1}$ which contradicts hypothesis \textbf{(H5)} as $\sigma'_{s_1(q_0)s_j(q_0)}$ is an off-diagonal and $\sigma'_{q_1}$ a diagonal term of $\Sigma'_q$. Therefore, for all $j \in [\![1 ; p]\!], s_j = s$ and $C'_a(.) = s(C_a(.))$. \\

Now, for $k_1, k_2 \in [\![1 ; q]\!]$ there exist a clustering $C_a$ and a couple of indices $j \neq l \in [\![1 ; p]\!]$ such that  $C_a(j) = k_1, C_a(l) = k_2$. Then, $\sigma_{k_1 k_2} = \Sigma_{a_{jl}} = \Sigma^{'}_{a_{jl}} = \sigma^{'}_{s(k_1) s(k_2)}$. Therefore $\Sigma' = \Sigma^{(s)}$ and $\Sigma'$ is equal to $\Sigma$ up to label permutation. \\

In particular, $C_a \Sigma C_a^{\top} = C^{'}_a \Sigma^{'}_q C^{'}_{a^{\top}}$ and thus $D' = \Sigma'_a - C'_a \Sigma C_a^{\top} = \Sigma_a - C_a \Sigma C_a^{\top} = D$.

For $\alpha$, the \citep{yakowitz1968identifiability} results prove that: $\forall k \in [\![1 ; q]\!], \alpha^{'p}_{k} = \alpha_{s(k)}^p$, with all $\alpha_k > 0$ so that $\forall k \in [\![1 ; q]\!], \alpha'_k = \alpha_{s(k)}$ and $\alpha' = \alpha^{(s)}$. \\

To conclude, under hypothesis \textbf{(H1)} to \textbf{(H5)}, there exists a permutation $s$ of $[\![1 ; q]\!]$ such that $(B', \Sigma^{'}_q, D', \alpha') = (B, \Sigma^{(s)}, D, \alpha^{(s)})$. This proves the model's identifiability up to label permutations.

\section{Concavity results}\label{appendix_concavity}
 For the various models the concavity proofs are based on the Hessian's (denoted $\mathcal{H}$) computation and analysis. 

\subsection{observed clusters spherical model}
 For this model we get the first-order differential:

\begin{align*}
    d J &= - \frac{np}{2} \xi d\xi^{-1} - \frac 12\mathrm{ tr}(R R^{\top}) d\xi^{-1} + \xi^{-1}\mathrm{ tr}(Y^{\top}XdB) - \xi^{-1}tr(B^{\top}X^{\top}XdB) \\
    & +\mathrm{ tr}(RC\mu^{\top})d\xi^{-1} - \xi^{-1}tr(C\mu^{\top}XdB) - \frac n2\mathrm{ tr}(C^{\top}C\Gamma)d\xi^{-1} - \frac 12\mathrm{ tr}(\mu(C^{\top}C)\mu^{\top}) d\xi^{-1}\\
    & + \frac n2\mathrm{ tr}(\Omega^{-1} d\Omega) - \frac n2\mathrm{ tr}(\Gamma d\Omega) - \frac 12\mathrm{ tr}(\mu^{\top} \mu d\Omega)
\end{align*}

 and the second-order differential:
\begin{align*}
    d^2 J &=  - \frac{np}{2}\xi^{2}d\xi^{-1}d\xi^{-1}  + 2\mathrm{ tr}(Y^{\top}XdB)d\xi^{-1} - 2\mathrm{ tr}(B^{\top} X^{\top}X dB)d\xi^{-1} \\
   & - \xi^{-1}tr(dB^{\top}X^{\top}XdB) - 2\mathrm{ tr}(C \mu^{\top}XdB)d\xi^{-1} - \frac n2\mathrm{ tr}(\Omega^{-1} d\Omega \Omega^{-1}d\Omega)
\end{align*}

 Hence the following hessian, where $\omega = \xi^{-1}$ and $R_{\mu} = (Y - X B - \mu C^T)$:
\begin{equation*}
    \begin{matrix}
    \partial^2\mathrm{vec}(\Omega) \\
    \partial^2\omega \\
     \partial^2\mathrm{vec}{(B})\\
    \end{matrix}
    \begin{pmatrix}
     -\frac n 2 \Sigma \otimes \Sigma & 0 & 0  \\
     & - \frac{np} 2 \omega^{-2} & 2 \mathrm{vec}\left(X^T R_\mu \right)\\
     & - \omega \left(I_p \otimes X^{\top} X\right)\\
    \end{pmatrix}
\end{equation*}

 We have $- \frac{np}{2} \omega^{-2} < 0$. $\Sigma$ is positive definite so that $-\frac n 2 \Sigma \otimes \Sigma $ is negative definite. $X^{\top} X$ is positive, it is positive definite if $X$ is full-rank so that $- \omega \left(I_p \otimes X^{\top} X\right)$ is negative, and negative definite if $X$ is full-rank, since $\omega > 0$. This proves the joint concavity of $J$ in $(\Omega, \xi^{-1})$ and in  $(\Omega, B)$.

\subsection{observed clusters general model}
 For this model we get the first-order differential:

\begin{align*}
    dJ &= \frac n2\mathrm{ tr}(D dD^{-1}) - \frac 12\mathrm{ tr}(R dD^{-1}R^{\top}) +\mathrm{ tr}(X dB D^{-1} Y^{\top}) -\mathrm{ tr}(D^{-1} B^{\top} X^{\top} X dB) \\
    & +\mathrm{ tr}(R dD^{-1} C \mu^{\top}) -\mathrm{ tr}(D^{-1} C \mu^{\top}XdB) - \frac n2\mathrm{ tr}(C \Gamma C^{\top} dD^{-1}) - \frac 12\mathrm{ tr}(\mu C^{\top} dD^{-1} C \mu^{\top}) \\
    & + \frac n2\mathrm{ tr}(\Sigma d\Omega) - \frac n2\mathrm{ tr}(d\Omega \Gamma) - \frac 12\mathrm{ tr}(\mu d\Omega d\mu^{\top})
\end{align*}

 and the second-order differential:
\begin{align*}
    d^2 J &= - \frac n2\mathrm{ tr}(D dD^{-1}D dD^{-1}) + 2\mathrm{ tr}(Y^{\top} X dB dD^{-1}) - 2\mathrm{ tr}(dD^{-1} B^{\top} X^{\top}X dB) \\
    & -\mathrm{ tr}(D^{-1} dB^{\top}X^{\top}XdB) - 2\mathrm{ tr}(dD^{-1} C \mu^{\top}XdB) - \frac n2\mathrm{ tr}(\Sigma d\Omega \Sigma d\Omega)
\end{align*}

 Hence the following hessian, where $R_{\mu} = (Y - X B - \mu C^T)$:
\begin{equation*}
    \begin{matrix}
    \partial^2\mathrm{vec}(\Omega) \\
    \partial^2 D^{-1} \\
     \partial^2\mathrm{vec}{(B})\\
    \end{matrix}
    \begin{pmatrix}
     -\frac n 2 \Sigma \otimes \Sigma & 0 & 0  \\
     & - \frac{n} 2 D \otimes D & 2 \mathrm{vec}\left(X^T R_\mu \right)\\
     & - D^{-1} \otimes X^{\top} X\\
    \end{pmatrix}
\end{equation*}

 As above, since $D$ is a diagonal matrix with strictly positive elements on the diagonal, we have that the diagonal terms are negative definite  hence the concavity results.

\subsection{Unobserved clusters model}
We recall that:
\begin{align*}
    J &= -n\left(\frac{p + q}{2}\right)\log(2\pi)  + \frac{nq}{2}\log(2 \pi e)  \\ 
    & - \frac{n}{2}\log(\det(D)) -\frac{1}{2}1_n^\top R^2  D^{-1} 1_p - \frac{1}{2} 1_n^\top M^{2} \tau^\top D^{-1}1_p - \frac{1}{2} 1_n^\top  S \tau^\top D^{-1} 1_p 
    \\ &+ 1_n^\top(R \odot  M \tau^\top )D^{-1}1_p
    \\ & + \frac{n}{2}\log\det(\Omega) - \frac{1}{2}1_n^\top(M \Omega \odot M)1_{q}  - \frac{1}{2}1_n^\top S \mathrm{diag}(\Omega) + \frac{1}{2}1_n^\top\log(S)1_{q} 
    \\ & + 1_p^\top\tau\log(\alpha) - 1_p^\top \left((\tau \odot \log(\tau)) \right)1_{q}
\end{align*}

\subsubsection{Concavity in $B$}
Let us consider only the terms of J whose double derivation in B is non-zero:
\begin{align*}
    \Tilde{J}(B) &= -\frac{1}{2}1_n^\top R^2  D^{-1} \\
    \partial^2_{B, B}J &= - \frac n2\mathrm{ tr}(D^{-1} dB^{\top} X^{\top} X dB ) \\
    \mathcal{H}_{B, B} &= - D^{-1} \otimes (X^{\top}X)
\end{align*}
$D^{-1}$ is a diagonal matrix with positive elements only, $X^{\top}X$ is always a positive matrix, and is definite positive if $X$ has full rank. Thus, $\mathcal{H}_{B, B}$ is a negative matrix, and $J$ is concave in $B$. 

\subsubsection{Concavity in $D^{-1}$}
Let us consider only the terms of J whose double derivation in $D^{-1}$ is non-zero:

\begin{align*}
    \Tilde{J}(D^{-1}) &= - \frac{n}{2}\log(\det(D)) \\
    \partial^2_{D^{-1}, D^{-1}}J &= - \frac{n}{2} \mathrm{tr}(D dD^{-1} D dD^{-1}) \\
    \mathcal{H}_{D^{-1}, D^{-1}} &= - \frac{n}{2} D \bigotimes D
\end{align*}
$D^{-1}$ is a diagonal matrix with positive elements only so that $\mathcal{H}_{D^{-1}, D^{-1}}$ is negative definite and $J$ is concave in $D^{-1}$

\subsubsection{Concavity in $\Omega$}
Let us consider only the terms of J whose double derivation in $\Omega$ is non-zero:

\begin{align*}
    \Tilde{J}(\Omega) &= \frac{n}{2}\log(\det(\Omega)) \\
    \partial^2_{\Omega, \Omega}J &=  - \frac n2 \mathrm{ tr}(\Sigma d\Omega \Sigma d\Omega)  \\
    \mathcal{H}_{\Omega, \Omega} &= -- \frac n2 \Sigma \otimes \Sigma
\end{align*}
$\Sigma$ is a positive definite matrix so that  only so that $\mathcal{H}_{\Omega, \Omega}$ is negative definite and $J$ is concave in $\Omega$

\subsubsection{Concavity in $\alpha$}

We can compute it "by hand" since $\alpha$ is a vector. 

\begin{align*}
    \tilde{J}(\alpha) &= 1_p^{\top} \tau \log(\alpha) \\
    \frac{\partial J}{ \partial \alpha_q} &= 
     \frac{\sum_{j=1}^p \tau_{jq}}{\alpha_q}
\end{align*}

$\mathcal{H}_{\alpha, \alpha}$ is a matrix of dimensions $q, q$ whose term $(k_1, k_2)$ is equal to $ \frac{\partial J}{ \partial \alpha_{k_1} \partial \alpha_{k_2}}$, which is equal to 0 if $k_1 \neq k_2$ and to $- \frac{\sum_{j=1}^p \tau_{jk_1}}{\alpha_{}^2}$ otherwise.

\subsubsection{Concavity in $M$}
Let us consider only the terms of J whose double derivation in $M$ is non-zero:
\begin{align*}
    \tilde{J}(M) &= - \frac 1 1_n^{\top} M^2 \tau^{\top} D^{-1} 1_p - \frac 12 1_n^{\top}(M \Omega \odot M) 1_{q} \\
    &= - \frac 12 \sum_{i,j} \sum_{k= 1}^{q} \tau_{jk}M_{ik}^2 D^{-1}_{jj} - \frac 12 \sum_{i = 1}^n \sum_{k_1, k_2 = 1}^{q} M_{ik_1}M_{ik_2}\Omega_{q_{k_1 k_2}} \\
    \frac{\partial  \tilde{J}}{\partial M_{ik_1}} &= - \sum_{j = 1}^p \tau_{jk_1} D^{-1}_{jj}  M_{ik_1} - \sum_{k_2 = 1}^{q} M_{ik_2} \Omega_{q_{k_1k_2}}\\
     \frac{\partial^2  \tilde{J}}{\partial^2 M_{ik_1}} &= - \sum_{j=1}^p \tau_{jk_1} D^{-1}_{jj} - \Omega_{q_{k_1k_1}} \\
     \frac{\partial^2  \tilde{J}}{\partial M_{ik_1}\partial M_{i k_2}} &= - \Omega_{q_{k_1k_1}} \text{ if } k_1 \neq k_2\\
     \frac{\partial^2  \tilde{J}}{\partial M_{i_1k_1}\partial M_{i_2k_2}} &= 0 \text{ if } i_1 \neq i_2 \text{ and } k_1 \neq k_2\\
     \mathcal{H}_{M, M} &= - I_n \otimes \left(\mathrm{diag} \left(\tau^{\top} D^{-1}1_p \right)-\Omega \right)
\end{align*}

$\tau$ only contains positive values and so does the diagonal of $D^{-1}$ so that  $\tau^{\top} D^{-1}1_p$ is positive definite and so is $\Omega$. Finally $\mathcal{H}_{M, M} = - I_n \otimes \left(\mathrm{diag} \left(\tau^{\top} D^{-1}1_p \right)-\Omega\right)$ is negative definite, we get the concavity in $M$.

\subsubsection{Concavity in $S$}
Considering only terms whose double derivation in $S$ is not equal to 0:
\begin{align*}
    \tilde{J}(S) &= \frac 12 \sum_{i, k} \log(S_{ik}) \\
    \frac{\partial  \tilde{J}}{\partial S_{ik}} &= \frac 12 \frac{1}{S_{ik}} \\
     \frac{\partial^2  \tilde{J}}{\partial^2 S_{ik}} &= -\frac 12 \frac{1}{S_{ik}^2} \\
     \frac{\partial^2  \tilde{J}}{\partial S_{i_1k_1}\partial S_{i_2 k_2}} &= 0 \text{ if } i_1 \neq i_2 \text{ or } k_1 \neq k_2\\
     \mathcal{H}_{S, S} &= - \frac 12 \mathrm{diag}\left(\mathrm{vec}\left(\frac{1}{S^2}\right)\right)
\end{align*}

\subsubsection{Concavity in $\tau$}
Considering only terms whose double derivation wrt $\tau$ is not equal to 0:
\begin{align*}
    \tilde{J}(\tau) &= - 1_p^{\top}\left(\tau \odot \log(\tau) \right) 1_{q} \\
    &= \sum_{j, k} \tau_{jk} \log( \tau_{jk}) \\
    \frac{\partial  \tilde{J}}{\partial \tau_{jk}} &= - \log(\tau_{jk}) - 1 \\
    \frac{\partial^2  \tilde{J}}{\partial^2 \tau_{jk}} &= - \frac{1}{\tau_{jk}}\\
    \frac{\partial^2  \tilde{J}}{\partial \tau_{j_1k_1} \partial \tau_{j_2k_2}} &= 0 \text{ if } j_1 \neq j_2 \text{ or } k_1 \neq k_2 \\
    \mathcal{H}_{\tau, \tau} &= - \mathrm{ diag }\left(\mathrm{vec}\left(\frac{1}{\tau}\right)\right)
\end{align*}

which is negative definite because $\tau$ only contains positive values. 
\section{EM criteria and estimators for the zero-inflated model}\label{appendix_zi_estimators}

We denote $ 0_Y = (1_{Y_{ij} = 0})_{i,j}, 1_Y = (1_{Y_{ij} \neq 0})_{i,j}$,  ${np}_Y = 1_n^\top 1_{Y} 1_p$, $n_Y = 1_n^\top 1_Y \in \mathbb{N}^p$, $R_{\mu} = Y - XB - \mu C^{\top}$. We also introduce the $n\times p$ matrix $\tilde{\Gamma}$, the rows of which are such that $\tilde{\Gamma}_i = \mathrm{diag}(C \Gamma^{(i)} C^\top) = (\Gamma^{(i)}_{q_j q_j})_{1 \leq j \leq p}$ for \emph{all} $j$, and $\Gamma_{row-sum} = \sum_{i=1}^n \Gamma_i $.
\subsection{Observed clusters}

For the zero-inflated Normal-Block, we obtain the following EM criterion:

\begin{align*}
J & = (0_Y \circ\delta_{0,\infty}(Y))_{total-sum} - \frac 12 \left({np}_Y + nq \right) \log(2\pi) \\
    & - \frac 12 n_{Y}^\top \log(d)  - \frac 12 \mathrm{tr}\left(D^{-1} 1_Y^\top R_\mu^2\right) - \frac 12 \mathrm{tr}\left(D^{-1} 1_Y^\top \tilde{\Gamma} \right) \\
    & + \frac n2 \log(\det(\Omega)) - \frac 12 \mathrm{tr}(\mu \Omega \mu^{\top}) -  \frac 12 \mathrm{tr}(\Omega \Gamma_{row-sum}) \\
    & + 1_n^\top \left(0_{Y}\log(\kappa) + 1_{Y}\log(1-\kappa) \right) 1_p\\
\end{align*}

One can also aqdd the entropy to retrieve the complete likelihood expression, at fixed parameters estimates:
\begin{align*}
     \hat{\ell}(\hat{B}, \hat{D}, \hat{\Sigma}_{q}, \hat{\kappa}) &= - \frac {{np}_Y} 2 \log(2\pi e) - \frac {n_Y^\top}{2} \log(\hat{d)}
-  \frac n 2 \log(\det(\hat{\Sigma}_{q})) + \frac 1 2 \sum_{i=1}^n \log |\hat{\Gamma}^{(i)}| \\
 & + 1_n \left(0_Y \log(\hat{\kappa}) + 1_Y \log(1 - \hat{\kappa})\right) - n \left(\kappa^\top \log(\kappa) + (1- \hat{\kappa})^\top \log(1 -\hat{\kappa})\right)
\end{align*}

 The E-step then consists in updating $\Gamma$ and $\mu$, the parameters of the posterior distributions $W_i | \mu_i$, for $i \in [\![ 1; n ]\!]$, for which we have explicit estimators. For the M-step, we update the estimates of $\Omega$, $\kappa$, $d$ and $B$. We have explicit estimators for the first three ones and need to use gradient descent to estimate $B$.

\begin{theorem}\label{zi_observed_estimators}
Below, the exponent $*_i$ indicates that we only consider $j$ for which $Y_{ij} \neq 0$. For the zero-inflated observed-clusters model, explicit estimators are given for $\mu, \Gamma, d, \kappa$ and $\Sigma$ by:
\begin{align*}
    \forall i \in [\![1 ; n ]\!], \Gamma^{(i)} 
    &= (\Omega + C^{*_i T} D^{*_i-1} C^{*_i})^{-1} \\
     \forall i \in [\![1 ; n ]\!], \mu^{(i)} &= \Gamma_i C^{*_i T}D^{*_i -1} (Y_i^{*_i} - B^{*_i T} X_i ) \\
     \Sigma &= \frac{1}{n} \left(\mu^{\top} \mu  +  \Gamma_{row-sum} \right)\\
     \kappa &= \frac{1}{n} 0_{\mathbf{Y}}^\top 1_n \\
     d &= \mathrm{diag}\left( 1_{Y}^\top \left(R^2_\mu + \tilde{\Gamma}\right)\right) \oslash n_Y
\end{align*}

$\hat{B}$ is estimated by maximizing $F(B) = - \frac 12 \mathrm{tr}\left(D^{-1} 1_Y^\top R_\mu^2\right)$ with $\nabla_B F(B) = X^\top \left(R_\mu D^{-1} \odot  1_Y \right)$
\end{theorem}

\subsection{Unobserved clusters}

 When the clustering is unobserved, we use a variational approximation, similarly to what is done for the non-zero-inflated model. Let $A=R^2 - 2 R \circ M \tau^T + (M^2+S)\tau^T$. For the ELBO, we have:

\begin{align*}
J & = (0_Y \circ\delta_{0,\infty}(Y))_{total-sum}   - \frac 12 \left({np}_Y  + nq \right) \log(2 \pi) + \frac{nq}{2} \log(2\pi e) \\
    & - \frac 12 1_n^{\top} \left(1_Y \odot A D^{-1} \right)1_p - \frac 12 {n}_Y^{\top} \log(d) \\
    & + \frac n2 \log(\det(\Omega))  - \frac 12 \text{tr}\left(\Omega (\text{diag}(S_{row-sum}) + M^{\top} M) \right) +\frac{1}{2} \log(S)_{total-sum} \\ 
    & +  1_n^\top \left(0_{Y}\log(\kappa) + 1_{Y}\log(1-\kappa) \right) 1_p + (\tau \log(\alpha))_{row-sum}  - (\tau \odot \log(\tau))_{total-sum}
\end{align*}

For the VE-step, we have explicit estimators for $S$ and $\tau$ but need to use a gradient descent for $M$. For the M-step, we have explicit estimators for $d$, $\Sigma$, $\alpha$ and $\kappa$, we use a gradient descent for $B$.

\begin{theorem}\label{zi_unobserved_estimators}
For the zero-inflated unobserved-clusters model, explicit estimators are given for $S$, $d$, $\Sigma$, $\alpha$ and $\kappa$ by:
\begin{align*}
    S &= \left(1_{Y}D^{-1}\tau + 1_n \mathrm{diag}(\Omega)\right)^{\oslash}\\
    d &= \mathrm{diag}\left( 1_{Y}^\top A \right) \oslash n_Y\\
    \Sigma &= \frac{1}{n}\left(M^\top  M + \mathrm{diag}(S^{T} 1_n)\right) \\
    \alpha &= \frac 1 n \tau_{row-sum} \\
    \kappa &= \frac 1n  0_{Y}^\top 1_n \\
    \forall j \in [\![ 1 ; p]\!], \tau_j &= \mathrm{softmax}(\eta_j) \\
    \mathrm{with} \text{  } & \eta = -\frac 12 D^{-1} \left(1_{Y}^\top (M^2 + S) -2 (1_{Y} \odot R)^\top M\right)+ 1_p \log(\alpha)^\top - 1_{qp}
\end{align*}

$M$ is estimated by maximizing $F(M) =  -\frac{1}{2} \left(1_n^\top \left(1_{Y}D^{-1} \odot \left(M^2 \tau^\top - 2 R \odot M \tau^\top \right)\right) 1_p  + 1_n^\top \left( (M \Omega \odot M) \right)1_{q}\right) $ with $\nabla_M F(M) = (1_Y D^{-1} \odot R) \tau - 1_Y D^{-1} \tau \odot M - M \Omega$.

B is estimated by maximizing $F(B) = -\frac{1}{2}1_n^\top\left(1_Y D^{-1}  \odot (R^2 - 2 R \odot M \tau^\top)\right) 1_p$ with $\nabla_B F(B) =  X^\top \left( 1_Y D^{-1} \odot (R - M \tau^{\top}) \right )$
\end{theorem}
\newpage
\section{ARI results for the Erdös-Rényi and Community network structures}\label{appendix_ari_res}

\begin{table}[h!]
    \centering
    \tiny
    \begin{tabular}{|p{0.3cm}|p{0.3cm}|p{0.3cm}|p{1.5cm}|p{1.5cm}|p{1.5cm}|}
        \hline
        n & p & q & Integrated inference - ARI mean (standard deviation) & 2-step method - variance clustering  - ARI mean (standard deviation) & 2-step method - residuals clustering - ARI mean (standard deviation)\\
        \hline
        20 & 100 & 3 & 1 (0) & 1 (0.02) & 1 (0)\\
        20 & 100 & 5 & 1 (0.01) & 0.96 (0.09) & 1 (0.01)\\
        20 & 100 & 10 & 0.97 (0.04) & 0.88 (0.09) & 0.98 (0.04)\\
        20 & 100 & 15 & 0.88 (0.08) & NA & 0.90 (0.06)\\
        20 & 500 & 3 & 1 (0) & 0.99 (0.02) & 1 (0)\\
        20 & 500 & 5 & 1 (0) & 0.98 (0.06) & 1 (0)\\
        20 & 500 & 10 & 0.99 (0.02) & 0.89 (0.10) & 0.99 (0.02)\\
        20 & 500 & 15 & 0.96 (0.03) & NA & 0.96 (0.04)\\
        50 & 100 & 3 & 1 (0) & 1 (0) & 1 (0)\\
        50 & 100 & 5 & 1 (0) & 1 (0) & 1 (0)\\
        50 & 100 & 10 & 1 (0) & 0.98 (0.04) & 0.98 (0.05)\\
        50 & 100 & 15 & 0.98 (0.11) & NA & 0.95 (0.11)\\
        50 & 500 & 3 & 1 (0) & 1 (0) & 1 (0)\\
        50 & 500 & 5 & 1 (0) & 1 (0) & 1 (0)\\
        50 & 500 & 10 & 1 (0) & 1 (0.01) & 0.98 (0.05)\\
        50 & 500 & 15 & 1 (0) & NA & 0.95 (0.04)\\
        200 & 100 & 3 & 1 (0) & 1 (0) & 1 (0)\\
        200 & 100 & 5 & 1 (0) & 1 (0) & 1 (0)\\
        200 & 100 & 10 & 1 (0) & 1 (0.01) & 0.98 (0.04)\\
        200 & 100 & 15 & 0.98 (0.13) & NA & 0.94 (0.11)\\
        200 & 500 & 3 & 1 (0) & 1 (0) & 1 (0)\\
        200 & 500 & 5 & 1 (0) & 1 (0) & 1 (0)\\
        200 & 500 & 10 & 1 (0) & 1 (0) & 0.97 (0.05)\\
        200 & 500 & 15 & 1 (0) & NA & 0.94 (0.03)\\
        500 & 100 & 3 & 1 (0) & 1 (0) & 1 (0)\\
        500 & 100 & 5 & 1 (0) & 1 (0) & 1 (0)\\
        500 & 100 & 10 & 1 (0) & 1 (0) & 0.97 (0.04)\\
        500 & 100 & 15 & 1 (0) & NA & 0.96 (0.03)\\
        500 & 500 & 3 & 1 (0) & 1 (0) & 1 (0)\\
        500 & 500 & 5 & 1 (0) & 1 (0) & 1 (0)\\
        500 & 500 & 10 & 1 (0) & 1 (0) & 0.97 (0.05)\\
        500 & 500 & 15 & 1 (0) & NA & 0.94 (0.03)\\
        \hline
    \end{tabular}
    \caption{ARI results for each configuration with the Erdös-Rényi network structure.}
    \label{tab:ari_res_ER}
\end{table}

\begin{table}[h!]
    \centering
    \tiny
    \begin{tabular}{|p{0.3cm}|p{0.3cm}|p{0.3cm}|p{1.5cm}|p{1.5cm}|p{1.5cm}|}
        \hline
        n & p & q & Integrated inference - ARI mean (standard deviation) & 2-step method - variance clustering  - ARI mean (standard deviation) & 2-step method - residuals clustering - ARI mean (standard deviation)\\
        \hline
        20 & 100 & 3 & 1 (0) & 1 (0) & 1 (0)\\
        20 & 100 & 5 & 1 (0.01) & 0.98 (0.04) & 1 (0.01)\\
        20 & 100 & 10 & 0.98 (0.03) & 0.90 (0.07) & 0.99 (0.02)\\
        20 & 100 & 15 & 0.92 (0.07) & NA & 0.93 (0.05)\\
        20 & 500 & 3 & 1 (0) & 1 (0.01) & 1 (0)\\
        20 & 500 & 5 & 1 (0) & 0.99 (0.03) & 1 (0)\\
        20 & 500 & 10 & 0.99 (0.02) & 0.92 (0.07) & 0.99 (0.02)\\
        20 & 500 & 15 & 0.98 (0.02) & NA & 0.97 (0.03)\\
        50 & 100 & 3 & 1 (0) & 1 (0) & 1 (0)\\
        50 & 100 & 5 & 1 (0) & 1 (0) & 1 (0)\\
        50 & 100 & 10 & 1 (0.01) & 1 (0.01) & 0.99 (0.03)\\
        50 & 100 & 15 & 1 (0.02) & NA & 0.97 (0.03)\\
        50 & 500 & 3 & 1 (0) & 1 (0) & 1 (0)\\
        50 & 500 & 5 & 1 (0) & 1 (0) & 1 (0)\\
        50 & 500 & 10 & 1 (0) & 1 (0) & 0.99 (0.03)\\
        50 & 500 & 15 & 1 (0) & NA & 0.96 (0.04)\\
        200 & 100 & 3 & 1 (0) & 1 (0) & 1 (0)\\
        200 & 100 & 5 & 1 (0) & 1 (0) & 1 (0)\\
        200 & 100 & 10 & 1 (0) & 1 (0) & 0.98 (0.04)\\
        200 & 100 & 15 & 1 (0) & NA & 0.96 (0.03)\\
        200 & 500 & 3 & 1 (0) & 1 (0) & 1 (0)\\
        200 & 500 & 5 & 1 (0) & 1 (0) & 1 (0)\\
        200 & 500 & 10 & 1 (0) & 1 (0) & 0.96 (0.06)\\
        200 & 500 & 15 & 1 (0) & NA & 0.93 (0.03)\\
        500 & 100 & 3 & 1 (0) & 1 (0) & 1 (0)\\
        500 & 100 & 5 & 1 (0) & 1 (0) & 1 (0)\\
        500 & 100 & 10 & 1 (0) & 1 (0) & 0.98 (0.04)\\
        500 & 100 & 15 & 1 (0) & NA & 0.96 (0.03)\\
        500 & 500 & 3 & 1 (0) & 1 (0) & 1 (0)\\
        500 & 500 & 5 & 1 (0) & 1 (0) & 1 (0)\\
        500 & 500 & 10 & 1 (0) & 1 (0) & 0.96 (0.05)\\
        500 & 500 & 15 & 1 (0) & NA & 0.93 (0.03)\\
        \hline
    \end{tabular}
    \caption{ARI results for each configuration with the community network structure.}
    \label{tab:ari_res_C}
\end{table}



\newpage
\bibliographystyle{elsarticle-harv} 
\bibliography{bibliography.bib}

\end{document}